\documentclass[12pt]{article}
\pdfoutput=1
\usepackage{amsmath,epsfig,a4,cite,latexsym,color,amssymb,etoolbox}
\usepackage{mathrsfs}
\usepackage[squaren,mediumqspace]{SIunits}
\usepackage{graphicx}
\usepackage{caption,subcaption}
\DeclareGraphicsExtensions{.pdf,.png,.jpg,.eps}
\usepackage{hyperref}
\newcommand{\amuSUSY}{a_{\mu}^{\text{SUSY}}}
\newcommand{\amured}{a_{\mu}^{\text{red}}}
\newcommand{\Deltamured}{\Delta_\mu^{\text{red}}}
\newcommand{\tbeff}{\tan\beta^{\text{eff}}}
\newcommand{\WHLcha}{{\tilde W}{\tilde H}{\tilde \nu}}
\newcommand{\WHLneu}{{\tilde W}{\tilde H}{\tilde \mu}_{L}}
\newcommand{\WHL}{{\tilde W}{\tilde H}{\tilde L}}
\newcommand{\BHL}{{\tilde B}{\tilde H}{\tilde \mu}_{L}}
\newcommand{\BHR}{{\tilde B}{\tilde H}{\tilde \mu}_{R}}
\newcommand{\BLR}{{\tilde B}{\tilde \mu}_{L}{\tilde \mu}_{R}}
\def\lsim{\mathrel{\rlap{\lower4pt\hbox{$\sim$}}
    \raise2pt\hbox{$<$}}}                
\def\gsim{\mathrel{\rlap{\lower4pt\hbox{$\sim$}}
    \raise2pt\hbox{$>$}}}                
\newcommand{\Mone}{M_{1}}
\newcommand{\Mtwo}{M_{2}}
\newcommand{\ML}{m_{L}}
\newcommand{\MR}{m_{R}}
\newcommand{\MUE}{\mu}
\newcommand{\BR}{\mathrm{BR}}
\newcommand{\sign}{\mathop{\mathrm{sign}}}
\begin{document}
\begin{flushright}
FTUV--15--1069, IFIC--15--01
\end{flushright}
\vspace{3em}
\begin{center}
{\Large\bf\boldmath Large muon $(g-2)$ with TeV-scale SUSY masses for $\tan\beta\to\infty$}
\\
\vspace{3em}
{Markus Bach$^\text{a}$, Jae-hyeon Park$^\text{b}$,  Dominik St\"ockinger$^\text{a}$, \\
Hyejung St\"ockinger-Kim$^\text{a}$
}\\[2em]
{$^\text{a}$ \sl Institut f\"ur Kern- und Teilchenphysik,
TU Dresden, 01069 Dresden, Germany} \\
{$^\text{b}$ \sl Departament de F\'{i}sica Te\`{o}rica and IFIC,
Universitat de Val\`{e}ncia-CSIC,
46100, Burjassot, Spain}
\setcounter{footnote}{0}
\end{center}
\vspace{2ex}
\begin{abstract}
The muon anomalous magnetic moment $a_\mu$ is investigated in the MSSM
for $\tan\beta\to\infty$.  
This is an attractive example of radiative muon mass generation with
completely different qualitative parameter dependence compared to the MSSM with
the usual, finite $\tan\beta$. 
The observed, positive difference between the experimental and Standard Model values
can only be explained if there are mass splittings, such that bino contributions dominate over wino ones. 
The two most promising cases are characterized either by large Higgsino mass $\mu$
or by large left-handed smuon mass $m_L$. 
The required mass splittings and the resulting $\amuSUSY$ are studied in detail. 
It is shown that the current discrepancy in $a_\mu$ can be explained
even in cases where all SUSY masses are at the TeV scale. The paper
also presents useful analytical formulas, approximations for limiting
cases, and benchmark points.

\end{abstract}

\vspace{0.5cm}

\newpage
\section{Introduction}

The muon anomalous magnetic moment $a_\mu=(g-2)_\mu/2$ provides
a tantalizing hint for physics beyond the Standard Model (SM)\@. The
current discrepancy between experiment and theory\footnote{
The experimental result has been obtained at BNL \cite{Bennett:2006fi};
a fourfold improvement in precision is expected from the new
experiments \cite{Carey:2009zzb,Iinuma:2011zz}. For
reviews of the theory prediction see
Refs.~\cite{JegerlehnerNyffeler,Miller:2012opa}; recent theory progress has
been achieved on the QED
\cite{Kinoshita2012,Kataev:2012kn,SteinhauserQED}, electroweak
\cite{Gnendiger:2013pva}, and  hadronic contributions
\cite{Davier,HMNT,Benayoun:2012wc,JegerlehnerSzafron,Kurz:2014wya}; for specific
reviews of the hadronic light-by-light contributions and expected
future improvements see \cite{Blum:2013xva,Benayoun:2014tra,Colangelo:2014pva,Masjuan:2014rea}. 
The value quoted in the text is based on the hadronic contributions from Ref.~\cite{Davier}.} is
\begin{align}
a_\mu^{\text{exp}}-a_\mu^{\text{SM}} &=
(28.7\pm8.0)\times10^{-10}.
\label{eq:amuexp-amusm}
\end{align}
It can be explained by a variety of new physics models below or at the TeV scale. 
Generally, the new physics contributions $a_\mu^\text{NP}$ are strongly correlated
with the loop contributions to the muon mass, $\delta m_\mu^\text{NP}$,
and they are suppressed by two powers of the typical new physics mass scale $M_{\text{NP}}$ \cite{CzM,Miller:2012opa}. 
This relation can be written as 
\begin{align}
a_\mu^{\text{NP}}=C_\text{NP}\frac{m_\mu^2}{M_\text{NP}^2} ,
\end{align}
where $C_\text{NP}={\cal O}(\delta m_\mu^\text{NP}/m_\mu)$ is given by
the model-dependent relative contribution to the muon mass.

Of special interest are, therefore, models of radiative muon mass
generation: in these models, $\delta m_\mu^\text{NP}$ amounts to the
entire physical muon mass and $C_\text{NP}={\cal O}(1)$.
Excluding fine-tuning in the muon mass, these models yield the largest  $a_\mu^\text{NP}$,
compared to other ones with the same new physics scale $M_{\text{NP}}$.
As the estimate $a_\mu^\text{NP}={\cal O}(m_\mu^2/M_{\text{NP}}^2)$ holds,
the observed deviation can in principle be explained
by a new physics scale of the order $\unit{2}{\tera\electronvolt}$.

The general idea of radiative muon mass generation can be realized in the
renormalizable and calculable framework of the minimal supersymmetric standard model (MSSM)\@. 
As discussed below in Sec.~\ref{sec:model} there are two distinct possibilities,
one of which has already been studied in Refs.~\cite{Borzumati:1999sp,Crivellin:2010ty,Thalapillil:2014kya}. 
The possibility considered in the present paper is the limit
\begin{align}
\tan\beta\equiv\frac{v_u}{v_d}\to\infty
\label{eq:tan beta to infty}
\end{align}
with the up- and down-type Higgs vacuum expectation values $v_{u,d}$.
In this limit $v_d$, the muon mass, and all other down-type lepton and quark masses vanish at tree level.
The masses arise from finite loop diagrams generating non-holomorphic couplings
of down-type fermions to the ``wrong'' Higgs doublet $H_u$. 
These loop diagrams are also important for finite $\tan\beta$,
and they have been discussed extensively in the literature,
often in the context of $B$-physics~\cite{tanbe},
but also in the context of the muon magnetic moment~\cite{Marchetti:2008hw}. 
Because of these loop diagrams, the limit $\tan\beta\to\infty$ exists and is phenomenologically viable \cite{Dobrescu:2010mk,Altmannshofer:2010zt}.  

We consider $a_\mu$ in the MSSM for $\tan\beta\to\infty$ for two reasons. 
On the one hand, this scenario is a calculable realization of the generally interesting idea of radiative muon mass generation,
and the results can be indicative of the more general situation. 
In particular, it is no special case of the model-independent analysis of
Ref.~\cite{Freitas:2014pua}, where simplified models with only two relevant particle masses have been considered. 
We will see that all cases of interest here involve at least three relevant
masses at the TeV scale.

On the other hand, the $\tan\beta\to\infty$ limit opens up an intriguing area of supersymmetry (SUSY)
parameter space where $a_\mu$ behaves qualitatively very differently
from the standard MSSM case (with moderate $\tan\beta$). This
standard MSSM case is well known (for reviews see
\cite{MartinWells,Stockinger:2006zn,Cho:2011rk}; 
recent works are
\cite{Endo:2013bba,Endo:2013lva,Endo:2013xka,Fargnoli:2013zda,Fargnoli:2013zia,
  Evans:2012hg,Ibe:2012qu,Ibe:2013oha,Bhattacharyya:2013xma,Mohanty:2013soa,Gogoladze:2014cha, 
Badziak:2014kea,Kowalska:2015zja,Chakrabortty:2015ika,Wang:2015rli}),
and it requires SUSY masses in the few-hundred GeV range in order to
explain the deviation~\eqref{eq:amuexp-amusm}; the LHC experiments,
however, start to exclude parts of the relevant parameter space
\cite{Endo:2013bba,Calibbi:2015kja}. Thus it is well motivated to ask
whether SUSY can explain the deviation even if all SUSY masses are
much higher, at the TeV scale. The answer can be expected to be
provided by SUSY radiative muon mass scenarios such as the one
considered here.

Our approach is a low-energy phenomenological one, with the aim to
answer whether and how TeV-scale SUSY masses are compatible with the
deviation~\eqref{eq:amuexp-amusm} in the MSSM limit
$\tan\beta\to\infty$, also taking into account other experimental data
and the internal consistency of the theory. Nevertheless it is
important to briefly take a top-down perspective and review how
infinite $\tan\beta$ might result from a more fundamental theory. One
appealing possibility is an unbroken continuous R-symmetry, which
forbids the $B\mu$ soft breaking term. Such a symmetry has been used
in Ref.~\cite{Davies:2011mp} to construct a ``one Higgs doublet model'', in
which $v_d=0$. R-symmetric models such as this one however contain
non-MSSM degrees of freedom, and therefore our MSSM study does not
directly apply to them. A mechanism which can naturally generate
vanishing/small $B\mu$ and infinite/large $\tan\beta$ in the MSSM is
gauge-mediated SUSY breaking. In its pure form \cite{Abel:2009ve} it requires
$B\mu=0$ at the messenger scale, and a non-vanishing $B\mu$ is only
generated by renormalization-group running at lower scales. The
connection between gauge-mediation and the large-$\tan\beta$ MSSM  has
been studied in detail in Ref.~\cite{Altmannshofer:2010zt}, however
with several simplifying restrictions which allow for very large but not
infinite $\tan\beta$. The promising results explained below provide
motivation for further model building to actually 
realize infinite $\tan\beta$ in a more fundamental theory, but this is
beyond the scope of the present paper.

We now give a brief preview of the MSSM limit $\tan\beta\to\infty$ to set the stage for the remainder of the paper. 
In the simple case that all relevant SUSY masses are equal to
$M_\text{SUSY}$ and $\tan\beta$ is moderate, the SUSY contribution to
$a_\mu$ is approximately given by the one-loop diagrams as
\begin{align}
a_\mu^\text{SUSY,1L}\approx
13 \times
10^{-10} \,\sign(\mu) 
\tan\beta \left(\frac{\unit{100}{\giga\electronvolt}}{M_{\text{SUSY}}}\right)^2.
\label{amuMSUSY}
\end{align}
For large $\tan\beta$ it has already been shown in
Ref.~\cite{Marchetti:2008hw} that higher-order terms leading in
$\tan\beta$ can become important and change the
linear dependence. These higher-order terms can be resummed as
\begin{align}
a_\mu^\text{SUSY}=
\frac{a_\mu^\text{SUSY,1L}}{1+\Delta_\mu},
\end{align}
where, still in the case of equal SUSY masses, the self-energy leads to
\begin{align}
\Delta_\mu\approx
 -  0.0018\,\sign(\mu )  \tan\beta.
\label{DeltamuMSUSY}
\end{align}
Hence, in the desired limit of infinite $\tan\beta$ the SUSY
contribution becomes 
\begin{align}
a_\mu^\text{SUSY}=
\lim_{\tan\beta\to\infty}\frac{a_\mu^\text{SUSY,1L}}{\Delta_\mu}
\approx 
- 72 \times 10^{-10} \,
 \left(\frac{\unit{1}{\tera\electronvolt}}{M_{\text{SUSY}}}\right)^2,
\label{negativeamuSUSY}
\end{align}
demonstrating that the magnitude of the contribution is very large, even for $M_\text{SUSY}$ in the multi-TeV range.
The proportionality to $\tan\beta$ has been replaced by a constant behaviour,
and the dependence on $\sign(\mu)$ has disappeared. However, the sign is predicted to be wrong,
so such a scenario with infinite $\tan\beta$ and equal SUSY masses is definitely excluded.

If the SUSY masses are not all equal, the above approximations do not apply,
and in the remainder of the paper we will give a full investigation of the five-dimensional parameter space. 
In particular we will characterize the regions that lead to a positive SUSY contribution to $a_\mu$,
which agrees with the observed deviation.

In Sec.~\ref{sec:model} we define the model, provide analytical
results and useful approximation formulas, and briefly review constraints from other sectors.
Comprehensive numerical analyses and their physical interpretation for different
mass parameter regions are presented in Sec.~\ref{sec:num}. We also
discuss relevant constraints on the parameter space.
Further discussion and the conclusions are given in Sec.~\ref{sec:con}.  
In addition, we present relevant benchmark points and scan plots showing the
possible values of the lightest SUSY mass parameter for which the
discrepancy~\eqref{eq:amuexp-amusm} can be explained.

\section{\boldmath The limit \texorpdfstring{$\tan\beta\to\infty$}{tan(beta)->infinity}}
\label{sec:model}

\subsection{Analytical results and approximations}

The MSSM tree-level muon mass is given by
\begin{align}
m_\mu^\text{tree}=y_\mu v_d,
\end{align}
the product of the muon Yukawa coupling $y_\mu$ and the down-type
Higgs vacuum expectation value $v_d$. Radiative muon mass generation in the MSSM requires the 
tree-level mass to vanish. There are two generic 
possibilities: either $y_\mu=0$ or $v_d=0$.
In case $y_\mu=0$, the
muon mass can be generated from loop diagrams involving binos and
smuons and the non-standard soft supersymmetry breaking term
$A_\mu'$. The value of $a_\mu$ in this scenario has already been studied in
Refs.~\cite{Borzumati:1999sp,Crivellin:2010ty}, 
and the observed deviation in $a_\mu$ can indeed be explained for bino
and smuon masses around $\unit{1}{\tera\electronvolt}$.
A similar study has been performed with the holomorphic trilinear coupling
in Ref.~\cite{Thalapillil:2014kya}.

Here we consider the second possibility $v_d=0$, or equivalently
the limit of Eq.~\eqref{eq:tan beta to infty}.
In the following we explain
the relevant formulas governing the muon mass, Yukawa coupling, and
magnetic moment in this limit.

The physical muon (pole) mass $m_\mu$ is given by the on-shell muon self
energy\footnote{All the following formulas are given for the limit
$\tan\beta\to\infty$. The formulas valid exactly also for arbitrary
(small, large or infinite)
$\tan\beta$ can easily be reconstructed by replacing
$\Deltamured\to\frac{v_d}{v_u}+\Deltamured$, in
Eqs.~(\ref{eq:mmufromSE},\,\ref{amuSUSYratio},\,\ref{eq:additionalformulas}).}
\begin{align}
  m_\mu = -\Sigma_\mu^\text{MSSM}
\equiv y_\mu v_u\Deltamured.
\label{eq:mmufromSE}
\end{align}
The explicit result can be found in the Appendix.
This equation is used to determine the value of the muon Yukawa
coupling $y_\mu$. 

Eq.~\eqref{eq:mmufromSE} also defines the ``reduced'' self energy factor
$\Deltamured$ (which satisfies 
$\Delta_\mu=\tan\beta\,\Deltamured$ with the usual definition for
$\Delta_\mu$ \cite{Marchetti:2008hw} and which agrees with the
quantity $\epsilon_\mu$ of Ref.~\cite{Altmannshofer:2010zt}). The
definition highlights the physics behind the muon 
mass generation: chiral symmetry (under which left- and right-handed
muon transform with different phases) is broken by the Yukawa
coupling, and the muon mass is generated by loop-induced couplings of
the muon to the vacuum expectation value of the ``wrong'' Higgs doublet
$v_u$. Hence the factor $y_\mu v_u$ can be pulled out of all
contributions to the muon mass. (Another factor of the Higgsino mass
$\mu$ could also be pulled out of all contributions, because $\mu$ has
to appear in all couplings of the muon to $v_u$ due to Peccei-Quinn
symmetry.) 
It should be noted that the coupling constants, mixing matrices,
and mass eigenvalues entering the self energy partially depend on the
muon Yukawa coupling, so $\Deltamured$ has a small residual dependence
on $y_\mu$. 

The combination 
\begin{align}
 y_\mu v_u \equiv
m_\mu\tbeff
  \label{eq:tbeff}
\end{align}
is thus an important quantity. In the standard case with moderate
$\tan\beta$ it is strictly equal to $m_\mu^\text{tree} \tan\beta $ and 
also often identified with $m_\mu\tan\beta$, but the latter identification is
only possible if either
higher-order effects can be ignored or on-shell
renormalization is used. Here we have to distinguish $m_\mu$ from
$m_\mu^\text{tree}$ and $\tbeff$ from $\tan\beta$.

The physics behind the generation of $a_\mu$ is the same as for the
muon mass. To highlight this similarity we write
\begin{align}
\amuSUSY=
\frac{y_\mu v_u}{m_\mu}
\amured,
\label{eq:amured}
\end{align}
introducing another dimensionless ``reduced'' quantity $\amured$. In
the standard case, 
we simply have $\amuSUSY\approx\tan\beta\,\amured$. Again, $\amured$ has
only a small residual dependence on $y_\mu$, and its explicit
result can be found in the Appendix.

Combining the equations for the muon mass generation and the magnetic
moment, we obtain the MSSM prediction in the limit
$\tan\beta\to\infty$:
\begin{align}
\amuSUSY = \frac{\amured}{\Deltamured}.
\label{amuSUSYratio}
\end{align}
We record here further useful relations between the quantities
introduced so far:
\begin{align}
\amuSUSY &= \tbeff\amured,&
y_\mu &= \frac{m_\mu}{v_u\Deltamured},&
\tbeff &= \frac{1}{\Deltamured}\approx1650\,y_\mu.
\label{eq:additionalformulas}
\end{align}

In contrast to Eqs.~(\ref{amuMSUSY},\,\ref{DeltamuMSUSY})
from the Introduction, the exact result Eq.~\eqref{amuSUSYratio}
depends on five independent SUSY mass parameters: the Higgsino
mass $\mu$, the gaugino (bino and wino) masses $M_{1,2}$, and the
left- and right-handed smuon soft mass parameters $m_{L,R}$. 
The result has interesting symmetry
properties due to cancellations 
between numerator and denominator: on the one hand,
$\amuSUSY$ is invariant under a 
change of sign of $\mu$ and, on the other, it is invariant
under a simultaneous change of signs of $\Mone$ and $\Mtwo$. This constitutes a
fundamental difference compared to the MSSM with finite $\tan\beta$, where the
sign of $a_\mu^\text{SUSY}$ can directly be specified via the sign of $\mu$
as e.g.\ in Eq.~\eqref{amuMSUSY}.

\begin{figure}
  \centering
  \begin{subfigure}{0.3\textwidth}
    \centering \includegraphics{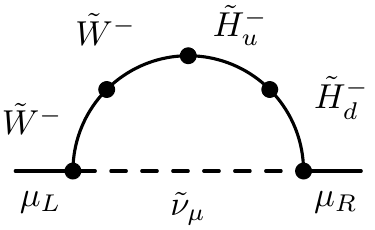}
    \caption{} \vspace{0.05\textwidth}
    \label{fig:HWsneu}
  \end{subfigure}
  \begin{subfigure}{0.3\textwidth}
    \centering \includegraphics{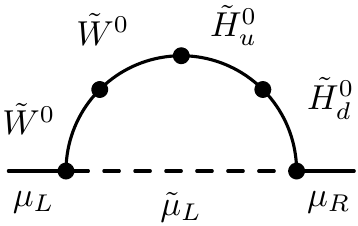}
    \caption{} \vspace{0.05\textwidth}
    \label{fig:HWL}
  \end{subfigure}
  \begin{subfigure}{0.3\textwidth}
    \centering \includegraphics{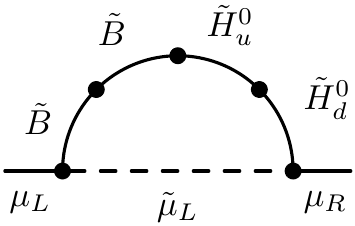}
    \caption{} \vspace{0.05\textwidth}
    \label{fig:HBL}
  \end{subfigure}
  \begin{subfigure}{0.3\textwidth}
    \centering \includegraphics{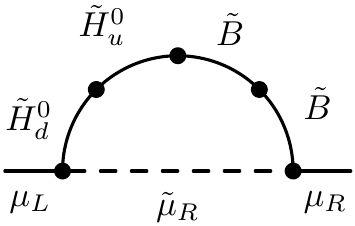}
    \caption{}
    \label{fig:HBR}
  \end{subfigure}
  \begin{subfigure}{0.3\textwidth}
    \centering \includegraphics{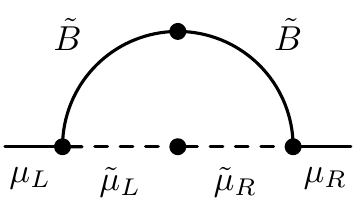}
    \caption{}
    \label{fig:BLR}
  \end{subfigure}
  \caption{Mass-insertion diagrams contributing to $\Sigma_\mu^\text{MSSM}$ and
  $\amuSUSY$. For the latter, an external photon couples to any of the
  charged particles in the loop.}
  \label{fig:massinsertion}
\end{figure}

Now, useful mass-insertion approximations for the MSSM muon self energy and the SUSY
contribution to the anomalous magnetic moment of the muon are
provided. Their validity for the parameter ranges of interest
will be discussed at the end of the subsequent section.
To maximize the magnitude of $a_\mu$ and minimize contraints from
$CP$-violating observables we restrict our considerations to real mass parameters $\mu$,
$M_{1,2}$. Taking advantage of the above-mentioned symmetries we choose,
without loss of generality, $\mu$ and $\Mone$ to be positive. Only the sign
of $\Mtwo$ remains arbitrary. Under the assumption $M_Z \ll M_\text{SUSY}$, the
SUSY contributions are given by the five mass-insertion
diagrams in Fig.~\ref{fig:massinsertion} \cite{moroi,Stockinger:2006zn,Cho:2011rk}.
The self energy factor $\Deltamured$ can then be decomposed as the sum of 
\begin{subequations}
\label{eq:sigmamuparts}
\begin{align}
  \Deltamured(\WHLcha)&=
    {-}\frac{g_2^2}{16 \pi^2} \,M_2 \mu \,I(M_2,\mu,m_L),\\
  \Deltamured(\WHLneu)&=
    {-}\frac{g_2^2}{32 \pi^2} \,M_2 \mu \,I(M_2,\mu,m_L),
    \label{eq:sigmamupartsHWL}\\
  \Deltamured(\BHL)&=
    \phantom{-}\frac{g_1^2}{32 \pi^2} \,M_1 \mu \,I(M_1,\mu,m_L),\\
  \Deltamured(\BHR)&=
    {-}\frac{g_1^2}{16 \pi^2} \,M_1 \mu \,I(M_1,\mu,m_R),\\
  \Deltamured(\BLR)&=
    \phantom{-}\frac{g_1^2}{16 \pi^2} \,M_1 \mu \,I(M_1,m_L,m_R),
\end{align}
\end{subequations}
where $g_{1,2}$ are the U(1) and SU(2) gauge couplings, 
and the loop function is given by
\begin{align}
  I(a,b,c)
  =\frac{a^2\,b^2\ln\frac{a^2}{b^2}+b^2\,c^2\ln\frac{b^2}{c^2}
  +c^2\,a^2\ln\frac{c^2}{a^2}}{\left(a^2-b^2\right)\left(b^2
  -c^2\right)\left(a^2-c^2\right)}.
\end{align}
This loop function is dimensionful and scales as $1/M^2$,
where $M$ denotes the largest of the three mass arguments.
The result for the self energy corresponds to the one from Ref.~\cite{Marchetti:2008hw} if the
limit $\tan\beta\to\infty$ is taken.

Likewise, for the case $M_Z \ll M_\text{SUSY}$, the SUSY contribution to
the anomalous magnetic moment of the muon can be approximated by the
sum of the expressions 
\begin{subequations}
\label{eq:amuparts}
\begin{align}
\amured(\WHLcha)&=
    \phantom{-} \frac{g_2^2}{\phantom{1}8 \pi^2} \, m_\mu^2 \frac{M_2 \mu}{m_L^4} \,
    F_a\left(\frac{M_2^2}{m_L^2},\frac{\mu^2}{m_L^2}\right),\\
  \amured(\WHLneu)&=
    -\frac{g_2^2}{16 \pi^2} \, m_\mu^2 \frac{M_2 \mu}{m_L^4} \,
    F_b\left(\frac{M_2^2}{m_L^2},\frac{\mu^2}{m_L^2}\right),
    \label{eq:amupartsHWL}\\
  \amured(\BHL)&=
    \phantom{-} \frac{g_1^2}{16 \pi^2} \, m_\mu^2 \frac{M_1 \mu}{m_L^4} \,
    F_b\left(\frac{M_1^2}{m_L^2},\frac{\mu^2}{m_L^2}\right),\\
  \amured(\BHR)&=
    - \frac{g_1^2}{\phantom{1}8 \pi^2} \, m_\mu^2 \frac{M_1 \mu}{m_R^4} \,
    F_b\left(\frac{M_1^2}{m_R^2},\frac{\mu^2}{m_R^2}\right),\\
  \amured(\BLR)&=
    \phantom{-} \frac{g_1^2}{\phantom{1}8 \pi^2} \, m_\mu^2 \frac{\mu}{M_1^3} \,
    F_b\left(\frac{m_L^2}{M_1^2},\frac{m_R^2}{M_1^2}\right),
\end{align}
\end{subequations}
corresponding to the mass-insertion diagrams in Fig.~\ref{fig:massinsertion},
with an external photon coupling to the charged internal line. Via
Eq.~\eqref{eq:amured}, this result can be related to the one for finite
$\tan\beta$ quoted e.g.\ in Refs.~\cite{Cho:2011rk,Fargnoli:2013zia},
where the loop functions 
\begin{subequations}
\begin{align}
  F_a(x,y)&=-\frac{F_2^C(x)-F_2^C(y)}{3 (x-y)},\\
  F_b(x,y)&=-\frac{F_2^N(x)-F_2^N(y)}{6 (x-y)}
\end{align}
\end{subequations}
have been introduced.
The functions $F_2^C$ and $F_2^N$ can be found in the Appendix.

We now provide seminumerical approximations which allow to directly
read off numerical orders of magnitude and signs and which facilitate
the phenomenological discussion. The dimensionless loop functions
\begin{subequations}
\begin{align}
  \hat I\left(\frac{a}{c},\frac{b}{c}\right)&=ab\,I(a,b,c),\\
  \hat K_N(x,y)&=2xy F_b(x,y),\\
  \hat K_W(x,y)&=2xy \big[2 F_a(x,y)-F_b(x,y)\big],
\end{align}
\end{subequations}
are especially useful if the sign and size of the different contributions shall be compared.
The first two of these functions have been introduced in Refs.~\cite{Dobrescu:2010mk}
and \cite{Endo:2013bba} respectively.
In the special case of three equal masses, we get $\hat I(1,1)=1/2$, 
$\hat K_W(1,1)=5/6$ and $\hat K_N(1,1)=1/6$. 
In general, $\hat I$ and $\hat K_N$ lie between $0$ and $1$; $\hat
K_W$ lies between $0$ and $1.155$. Furthermore, in the case of equal
arguments, $\hat I(x,x)$  and $\hat K_N(x,x)$ increase
monotonically, while $\hat K_W(x,x)$ increases up to a value of about 1.155 at
$x\approx 11.15$ and then decreases towards a limit of 1 for $x\to\infty$.

By inserting all numerically known Standard Model quantities we obtain
a seminumerical version of Eqs.~\eqref{eq:sigmamuparts} and the negative MSSM muon
self energy, which should be equal to the muon mass [see Eq.~\eqref{eq:mmufromSE}],
can be written as
\begin{align}
m_\mu &= y_\mu v_u\Deltamured \nonumber\\
&
\begin{aligned}\approx
  y_\mu\bigg[
  -\sign(M_2)\,&\unit{0.705}{\giga\electronvolt}\,
  \hat I\left(\frac{|M_2|}{m_L},\frac{\mu}{m_L}\right)\\
  {}+{}&\unit{0.067}{\giga\electronvolt}\,
  \hat I\left(\frac{M_1}{m_L},\frac{\mu}{m_L}\right)\\
  {}-{}&\unit{0.135}{\giga\electronvolt}\,
  \hat I\left(\frac{M_1}{m_R},\frac{\mu}{m_R}\right)\\
  {}+{}&\unit{0.135}{\giga\electronvolt}\,
  \hat I\left(\frac{M_1}{m_R},\frac{m_L}{m_R}\right)\frac{\mu}{m_L}
  \bigg],
\end{aligned} \label{eq:sigmamuapprox}
\end{align}
where the order of terms is the same as in Eqs.~\eqref{eq:sigmamuparts}, and the first
term combines the contributions from loops with charged or neutral winos.

Similarly, by plugging in numbers into Eqs.~\eqref{eq:amuparts}, we
obtain the
seminumerical mass-insertion approximation for $\amuSUSY$ in the
$\tan\beta\to \infty$ limit as 
\begin{align}
  \amuSUSY 
  &= \frac{y_\mu v_u}{m_\mu} \amured \nonumber\\
  &
  \begin{aligned}\approx
  y_\mu \bigg[
  \sign(M_2)\, &2.483 \times 10^{-8}\,
  \frac{\left(\unit{1}{\tera\electronvolt}\right)^2}{\mu |M_2|}\,
  \hat K_W \left(\frac{M_2^2}{m_L^2},\frac{\mu^2}{m_L^2}\right) \\
  +\,&0.712\times 10^{-8}\,
  \frac{\left(\unit{1}{\tera\electronvolt}\right)^2}{\mu M_1}\,
  \hat K_N \left(\frac{M_1^2}{m_L^2},\frac{\mu^2}{m_L^2}\right) \\
  -\,&1.425\times 10^{-8}\,
  \frac{\left(\unit{1}{\tera\electronvolt}\right)^2}{\mu M_1}\,
  \hat K_N \left(\frac{M_1^2}{m_R^2},\frac{\mu^2}{m_R^2}\right) \\
  +\,&1.425\times 10^{-8}\,
  \frac{\mu M_1\left(\unit{1}{\tera\electronvolt}\right)^2}{m_L^2 m_R^2}\,
  \hat K_N \left(\frac{m_L^2}{M_1^2},\frac{m_R^2}{M_1^2}\right)
   \bigg],
  \end{aligned}
  \label{eq:amuapprox}
\end{align}
which corresponds to the result from Ref.~\cite{Endo:2013bba} in the limit $\tan\beta\to\infty$ and combines
the contributions that are created by a charged or neutral internal wino in the
first term.

The approximate expressions~\eqref{eq:sigmamuapprox} and \eqref{eq:amuapprox}
have several noteworthy features.
\begin{itemize}
\item They are linear in $y_\mu$ because the residual dependence of
  $\Deltamured$ and $\amured$ on $y_\mu$ vanishes for
  $M_Z/M_\text{SUSY}\to0$.
\item All terms
   involve the factor $\mu M_1$ or $\mu M_2$. This
  explains the symmetry of $\amuSUSY$ (given by Eq.~\eqref{amuSUSYratio})
under a sign change either of $\mu$ or
  of $M_1$ and $M_2$.
\item Under sign change of $M_2$ alone, the wino contributions change
  their signs relative to the bino contributions.
\item The signs of the contributions are related: each
  mass-insertion diagram contributes with equal sign to $\Deltamured$
  and $\amured$, except for the chargino diagram, where the signs are
  opposite.
\item Thus if the chargino diagram
(\subref{fig:HWsneu}) in Fig.~\ref{fig:massinsertion} dominates in both
$\Deltamured$ and $\amured$, we
get $\amuSUSY<0$. In contrast, for neutralino dominance
$\amuSUSY$ becomes positive and, therefore, has the
correct sign with regard to an explanation of the
discrepancy~\eqref{eq:amuexp-amusm}.
\item The
contributions~\eqref{eq:sigmamupartsHWL} and \eqref{eq:amupartsHWL} from the
diagram with an internal neutral wino can never dominate since they are always
smaller than the ones from the diagram with an internal charged
wino. In $\Deltamured$, the charged and neutral wino contribution add
constructively, in $\amured$ destructively.
\end{itemize}

\subsection{Experimental constraints from other sectors}
\label{sec:expconstraints}

Before we present our numerical analyses,  
we briefly discuss constraints on parameter space from other
observables and sectors and show that the arising constraints can be
satisfied without restricting the five parameters
$\mu,M_{1,2},m_{L,R}$ relevant for $a_\mu$. Clearly, an observable
strongly related to $a_\mu$ is 
the lepton flavour violating decay $\mu\to e\gamma$. The correlation
of the two MSSM predictions has been discussed extensively in the literature
\cite{Graesser:2001ec,Chacko:2001xd,Isidori:2007jw,Kersten:2014xaa,Chiu:2014oma}.
If the prediction for $\amuSUSY$ is fixed, the SUSY prediction
for $\mu\to e\gamma$
can be estimated quite well up to the unknown lepton flavour violating
parameters. In particular the one-loop amplitudes for the two
quantities share the same $\tan\beta$ enhancement; the
correlation between the two does not strongly depend on
$\tan\beta$. E.g.\ Figs.~13--16 of Ref.~\cite{Kersten:2014xaa}
remain  valid also for $\tan\beta\to\infty$. Therefore, like for
moderate $\tan\beta$, the decay $\mu\to
e\gamma$ is compatible with experimental bounds \cite{Adam:2013mnn}
for sufficiently small lepton flavour violating parameters.

Further constraints from $B$-physics have been discussed extensively
in Ref.~\cite{Altmannshofer:2010zt}; taking the limits of those results
as $\tan\beta\to\infty$ shows that
agreement between theory and experiment can be achieved. The
parameters mainly constrained by $B$-physics are the heavy Higgs
boson masses as well as the stop trilinear coupling $A_t$.
Since these parameters are not
relevant for the discussion of $a_\mu$, we give only a brief account of
how they are constrained.
The new physics contribution to
$\BR(B^+ \rightarrow \tau^+ \nu)$ can be suppressed below $2 \sigma$
with $\sigma$ being the experimental error,
by having the charged Higgs mass $M_{H^+}$ around a few TeV or higher.
We can relax the constraints from
$B_s \rightarrow \mu^+ \mu^-$ and
$B \rightarrow X_s \gamma$
by assuming that $A_t$ and flavour-violating squark mass insertions are
small enough.
In addition,
the former process can be further suppressed by
raising the masses of the $CP$-odd as well as the $CP$-even heavier Higgses,
and the latter by
increasing the squark masses.

Given large bottom and tau Yukawa couplings, large 
sbottom and stau corrections
to the mass of the lighter $CP$-even Higgs $h$ might undermine
the stop contribution that should lift $m_h$ up to the measured value
around $\unit{125}{\giga\electronvolt}$ \cite{HiggsMassLift}, as $\tan\beta\to\infty$.
Indeed, Fig.~2 of Ref.~\cite{Brignole:2002bz} shows a rapid drop
of $m_h$ as $\tan\beta$ grows over around 40.
This happens due to a cancellation between the tree-level and the
$\Delta_b$ contributions to the bottom quark pole mass causing
a blow-up of the bottom Yukawa coupling.
As mentioned in the same reference,
this is however not necessarily the case for a high $\tan\beta$.
First of all, it never happens for $\mu < 0$.
Even if $\mu > 0$, once we enter
the regime where $|\Delta_b| \gg 1$, the bottom Yukawa coupling
comes back to $\mathcal{O}(1)$, thereby suppressing its negative contribution to $m_h$.
Therefore the $\tan\beta\to\infty$ scenario allows to choose third generation sfermion masses
which lead to a correct Higgs mass.

Owing to decoupling of the multi-TeV extra Higgs masses
required by the $B$-physics constraints, the MSSM Higgs sector is approximately SM-like \cite{Gunion:2002zf}.
Nonetheless it is instructive to qualitatively discuss the most relevant SM-like
Higgs decay modes,
$h \rightarrow b \overline{b}$,
$h \rightarrow \tau^+ \tau^-$, and
$h \rightarrow \mu^+ \mu^-$,
which might potentially be affected
by large bottom and charged lepton Yukawa couplings if the heavy Higgs
masses are not sufficiently high to apply the decoupling limit.
Their decay rates can be kept close to the SM values
in the following way.
The effective $H_u^0$-$f$-$f$ coupling,
for $f = b, \tau, \mu$,
is generated by the same loop diagrams that generate $m_f$.
Therefore, the $h$-$f$-$f$ coupling should be SM-like if
$h$ consists purely of $H_u^0$ to a sufficient extent.
Remember that the $H_d^0$-$f$-$f$ coupling, $y_f$,
can be comparable to or even larger than $y_t$.
The remaining question is then how to maintain the Higgs mixing angle $\alpha$
small enough for $h$ to avoid excessive $H_d^0$-contamination.
It is easy to see that
the $2 \times 2$ tree-level $CP$-even Higgs mass matrix becomes diagonal
as $\tan\beta\to\infty$ (see e.g.\ Eq.~(23) of Ref.~\cite{Degrassi:2001yf}).
The one-loop corrections to its off-diagonal elements are suppressed
for small $A_t$ and $A_b$, which can follow from gauge-mediated SUSY
breaking at the same time as the small $B\mu$, needed for large $\tan\beta$.
Regarding $h\rightarrow\gamma\gamma$, we can use the fact that
the stau-loop contribution
(see e.g.\ Refs.~\cite{Kitahara:2012pb,Carena:2012mw,Kitahara:2013lfa})
to the dimension-five effective operator decouples
faster than the tau self energy, as the stau masses increase.

\section{Numerical analysis} 
\label{sec:num}

\subsection{Dominance transition and sign change}

In the Introduction we have shown that if all SUSY masses are equal,
$\amuSUSY$ becomes negative in the limit of $\tan\beta\to\infty$. 
In this case the wino contributions dominate in both $\amured$ and $\Deltamured$,
i.e.\ in the numerator and denominator of Eq.~\eqref{amuSUSYratio},
and they have opposite signs, see Eqs.~(\ref{eq:sigmamuapprox},\,\ref{eq:amuapprox}). 
However, these equations also show that  the resulting $\amuSUSY$ will be positive
if any of the three bino contributions dominates. 

In the following we will delineate regions in the five-dimensional parameter space
of the masses $M_{1,2}$, $\mu$, $m_{L,R}$ in which $\amuSUSY$ is positive.
We start by noting that the signs only depend on mass ratios and that each of the
three bino contributions can be expected to dominate if a particular mass hierarchy is valid:
\begin{itemize}
\item The $\BLR$ contributions dominate for $\Mone, \ML, \MR \ll \mu$: ``large $\mu$-limit''
\item The $\BHR$ contributions dominate for $\Mone, \mu, \MR \ll \ML$: ``$\tilde{\mu}_R$-dominance''
\item The $\BHL$ contributions dominate for $\Mone, \mu, \ML \ll |\Mtwo|, \MR$.
These have the smallest numerical prefactors, and we can expect them to dominate only
for rather extreme hierarchies.\footnote{Dominance of the $\BHL$ contributions can also
be achieved with $|\Mtwo| \ll \Mone, \mu, \ML \ll \MR$.
We will not investigate this below since the mass hierarchies would be even more extreme.}
\end{itemize}
The phenomenological behaviour and importance of the $\BLR$ and $\BHR$ contributions
have also been discussed recently in various contexts in
Refs.~\cite{Endo:2013xka,Endo:2013lva,Endo:2013bba,Fargnoli:2013zda,Fargnoli:2013zia,Kersten:2014xaa,Badziak:2014kea}
and Refs.~\cite{Grothaus:2012js,Fargnoli:2013zda,Fargnoli:2013zia,Kersten:2014xaa}, respectively.
We first focus on cases where these two contributions become dominant. 
Parameter regions in which the $\BHL$ contributions dominate will be presented later on.

\begin{figure}[t]
\begin{subfigure}[b]{0.5\textwidth}
\includegraphics[scale=1]{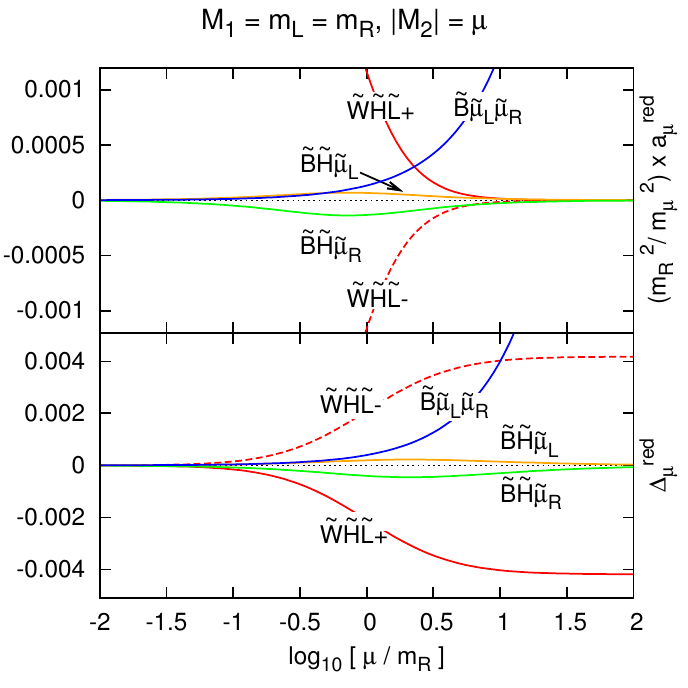}
\caption{large $\mu$-limit}
\label{fig:ay0}
\end{subfigure}
\begin{subfigure}[b]{0.5\textwidth}
\includegraphics[scale=1]{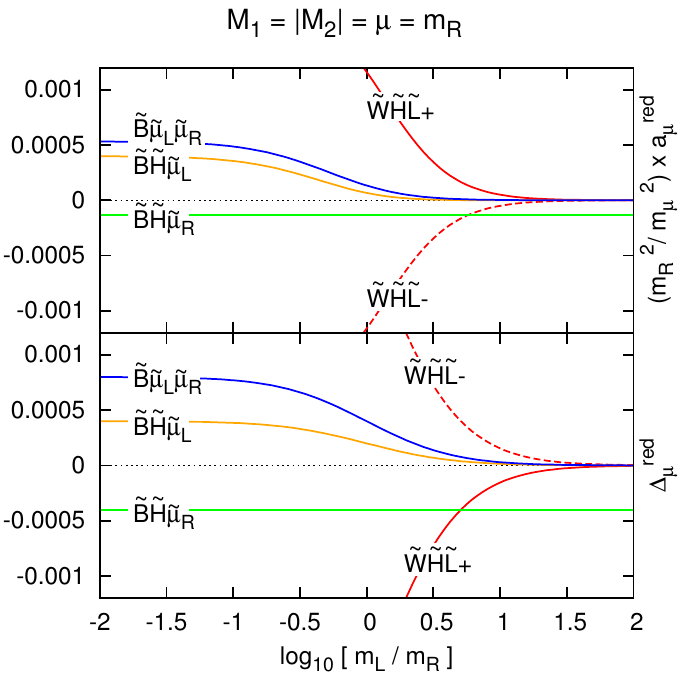}
\caption{$\tilde{\mu}_{R}$-dominance}
\label{fig:ax0}
\end{subfigure}
\caption{Individual behaviour of the contributions to $\amured$ and $\Deltamured$ for two different mass spectra. 
The full result $\amuSUSY$ is given by Eq.~\eqref{amuSUSYratio} as the ratio of the sums of the contributions. 
$\WHL\pm$ denotes the sum of $\WHLcha$ and $\WHLneu$, for positive or negative $\Mtwo$, respectively. 
In the upper plots, the individual contributions to $\amured$ are rescaled by $m_{R}^{2}/m_{\mu}^{2}$
to make the results depend only on ratios of SUSY masses.
}
\label{fig:fivelines}
\end{figure}
Fig.~\ref{fig:fivelines} illustrates plots of the individual contributions to $\amured$ and $\Deltamured$
for parameters which interpolate between the equal SUSY mass case and the two mass hierarchies
for either ``large $\mu$-limit'', Fig.~\ref{fig:fivelines}(\subref{fig:ay0}),
or ``$\tilde{\mu}_R$-dominance'', Fig.~\ref{fig:fivelines}(\subref{fig:ax0}). 
In each hierarchy case the desired bino contributions become the largest ones for sufficiently large mass ratios.
We first discuss Fig.~\ref{fig:fivelines}(\subref{fig:ay0}), corresponding to the ``large $\mu$-limit'', in detail. 
It shows the contributions as functions of the ratio $|\Mtwo| = \mu$ over $M_1=m_L=m_R$;
the horizontal axis is given by $x=\log_{10}(\mu/m_R)$. 
We pay particular attention to the dominance transition among the contributions
and the sign change of $\amured$ and $\Deltamured$. 
\begin{itemize}
\item Around  $x = 0$, where all masses are equal, the wino contribution dominates in both $\amured$ and $\Deltamured$.
For larger values of $x$, the $\BLR$ contributions increase proportionally to $\mu$ whereas the other contributions
are suppressed as $1/\mu$. Hence the latter tend to zero for large $x$, except for
the $\WHL$ contribution to $\Deltamured$, which tends to a constant
because it is also proportional to $\Mtwo$ and $|\Mtwo| = \mu$ in this plot. 
\item At $x \approx 0.36$, the dominant contribution to $\amured$ changes from the wino contribution
to the $\BLR$ contribution. For larger mass ratios, $x \approx 1.00$, the same dominance change also happens
in $\Deltamured$ and thus $\amuSUSY$ is always positive for $x>1$.
\item In the intermediate region, $0.36<x<1.00$, $\BLR$ already dominates in the numerator
but the wino contribution still dominates in the denominator of $\amuSUSY$. 
Here the sign of $\amuSUSY$ depends on $\Mtwo$ in such a way that it is equal to $\sign(-M_2)$.
\item For $x < 0$, the wino contributions remain dominant. 
\end{itemize}

Similarly, we discuss Fig.~\ref{fig:fivelines}(\subref{fig:ax0}), which corresponds to ``$\tilde{\mu}_R$-dominance''.
It shows the contributions as functions of the ratio $m_L$ over $M_1 = |\Mtwo| = \mu = m_R$;
the horizontal axis is given by $x = \log_{10}(m_L/m_R)$. 
\begin{itemize}
\item
Like in Fig.~\ref{fig:fivelines}(\subref{fig:ay0}), the equal mass case emerges at $x = 0$,
and around this point the wino contributions dominate in both the numerator and denominator of $\amuSUSY$.
For larger $x$, all contributions tend to zero except for the $\BHR$ ones,
which are independent of $m_L$ and thus stay constant. 
\item
At $x \approx 0.7$, the $\BHR$ contributions become dominant;
accidentally, the dominance change occurs approximately at the same $x$ in the numerator and denominator. 
For larger values of $x$, this leads to a positive $\amuSUSY$, independently of the sign of $\Mtwo$.
\end{itemize}

\begin{figure}
\begin{subfigure}[b]{\textwidth}
\begin{subfigure}[b]{0.5\textwidth}
\includegraphics[scale=1.0]{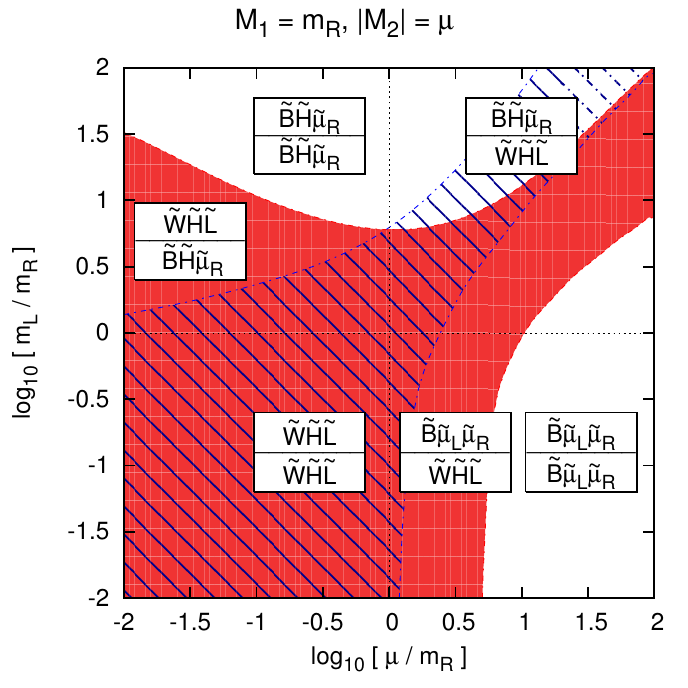}
\caption{}
\label{fig:signs1}
\end{subfigure}
\begin{subfigure}[b]{0.5\textwidth}
\includegraphics[scale=1.0]{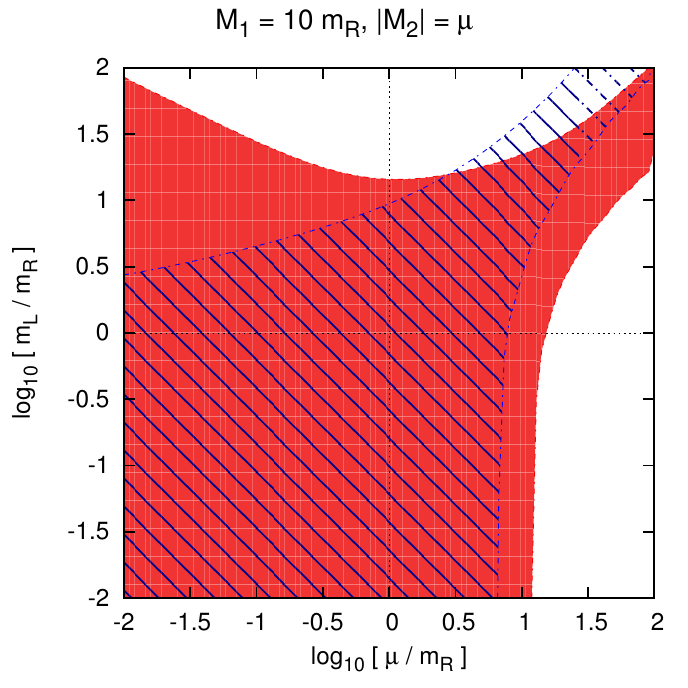}
\caption{}
\label{fig:signs2}
\end{subfigure}\vspace{0.03\textwidth}
\end{subfigure}
\begin{subfigure}[b]{\textwidth}
\begin{subfigure}[b]{0.5\textwidth}
\includegraphics[scale=1.0]{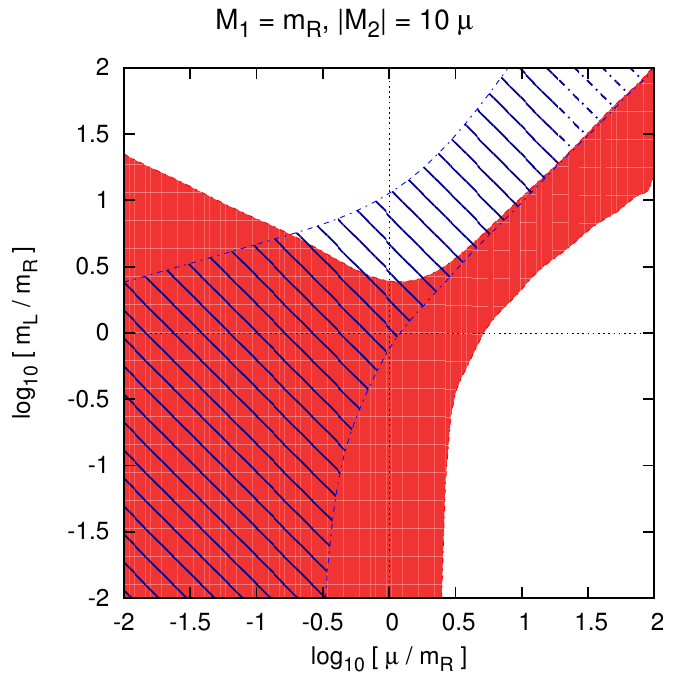}
\caption{}
\label{fig:signs3}
\end{subfigure}
\begin{subfigure}[b]{0.5\textwidth}
\includegraphics[scale=1.0]{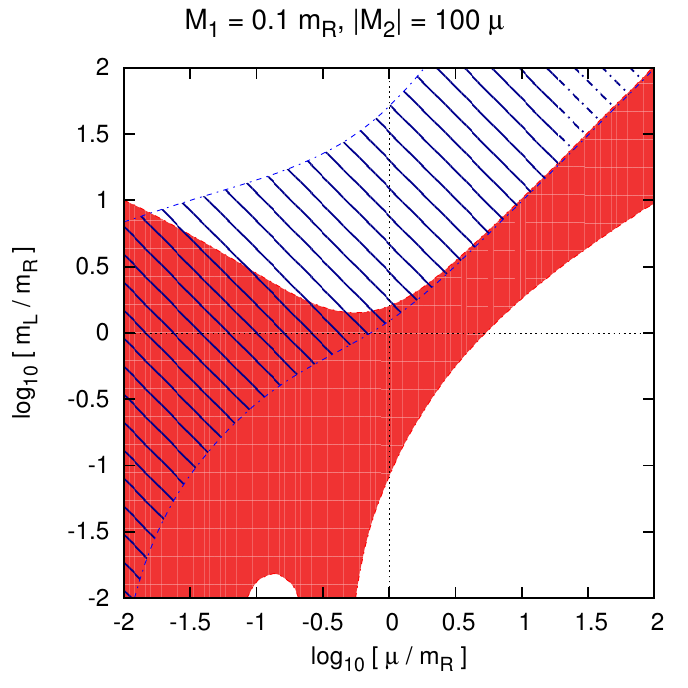}
\caption{}
\label{fig:signs4}
\end{subfigure}
\end{subfigure}
\caption{Signs of $\amuSUSY$ in the plane of $\ML/\MR$ versus $\MUE/\MR$, for the two possible signs of $\Mtwo$. 
In the white regions $\amuSUSY$ is positive. 
The red regions indicate negative $\amuSUSY$ for $\Mtwo>0$, the blue hatched ones negative $\amuSUSY$ for $\Mtwo<0$.
In the overlap regions $\amuSUSY$ is always negative. 
The first plot also displays which contributions dominate in the bulk of each region
in the numerator and denominator of Eq.~\eqref{amuSUSYratio}.
(It should be noted that near the borders between the regions there are cancellations,
and different contributions can be the largest.)
In each plot the remaining two mass ratios $\Mone/\MR$ and $|\Mtwo|/\MUE$ are fixed as indicated;
the structure of the dominance regions is the same in all plots.
For the small white region in the bottom left area of plot (d), see the text.} 
\label{fig:signs}
\end{figure}

To investigate the sign of $\amuSUSY$ comprehensively we plot it in a more general parameter space,
as a function of the two mass ratios used in Figs.~\ref{fig:fivelines}(\subref{fig:ax0}) and (\subref{fig:ay0}),
i.e.\ in the plane $\ML/\MR$ versus $\MUE/\MR$.  
Fig.~\ref{fig:signs} displays the results for four different choices
of the remaining mass ratios $\Mone/\MR$ and $|\Mtwo|/\MUE$.
We begin by discussing Fig.~\ref{fig:signs}(\subref{fig:signs1}).
It generalizes Fig.~\ref{fig:fivelines} by assuming $\Mone=\MR$ and $|\Mtwo|=\mu$,
so the different regions in Fig.~\ref{fig:signs}(\subref{fig:signs1}) can
be fully understood from the previous discussion. 
\begin{itemize}
\item The large, central red/blue overlap region contains the origin at which all relevant SUSY masses are equal. 
In this region the wino contributions dominate and $\amuSUSY$ is always negative.
\item The two white regions, where $\amuSUSY$ is always positive,
correspond to the ``large $\mu$-limit'' and the ``$\tilde{\mu}_R$-dominance'' region.
\item The blue hatched and the red regions (excluding the red/blue
  overlap) are transition regions where different contributions
  dominate in numerator and denominator. They generalize the region
  $0.36<x<1.00$ of Fig.~\ref{fig:fivelines}(\subref{fig:ay0}).
  Here the sign of $\amuSUSY$ depends on $\Mtwo$ and is equal to the sign of either
  $(+\Mtwo)$ (blue hatched) or $(-\Mtwo)$ (red). 
\end{itemize}

In Fig.~\ref{fig:signs}(\subref{fig:signs1}), the gaugino masses
$|\Mtwo|$ and $\Mone$ are equated with $\mu$ and $\MR$, respectively.  
This is a reasonable choice since their precise values are qualitatively unimportant
for the appearance of the different regions.  
The dependence on these two gaugino masses is investigated in
Figs.~\ref{fig:signs}(\subref{fig:signs2})--(\subref{fig:signs4}), 
whose presentation is similar to Fig.~\ref{fig:signs}(\subref{fig:signs1})
but with either $M_1 = 10\, m_R$, $|M_2| = 10\, \mu$, or $M_1 = 0.1\, m_R$ and $|M_2| = 100\, \mu$. 
Together, these plots cover large, representative parts of the full parameter space of the four relevant mass ratios.  

Obviously, all plots can be understood in the same way as Fig.~\ref{fig:signs}(\subref{fig:signs1}).
As $\Mone$ becomes heavier than $m_{R}$ in Fig.~\ref{fig:signs}(\subref{fig:signs2}),
the bino contributions are suppressed, the wino-dominance area expands, and the positive $\amuSUSY$ regions shrink.
In contrast, the positive $\amuSUSY$ regions expand in Figs.~\ref{fig:signs}(\subref{fig:signs3})
and (\subref{fig:signs4}), since $\Mtwo$ gets heavier than $\mu$ and the wino contributions are thereby suppressed.

\begin{figure}
\begin{subfigure}{0.5\textwidth}
\includegraphics[scale=1]{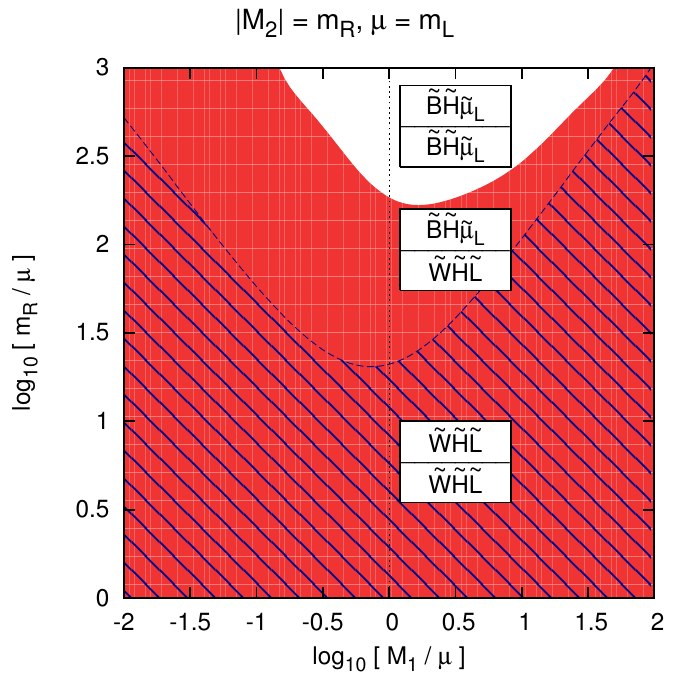}
\caption{}
\label{fig:signBHL}
\end{subfigure}
\begin{subfigure}{0.5\textwidth}
\includegraphics[scale=1]{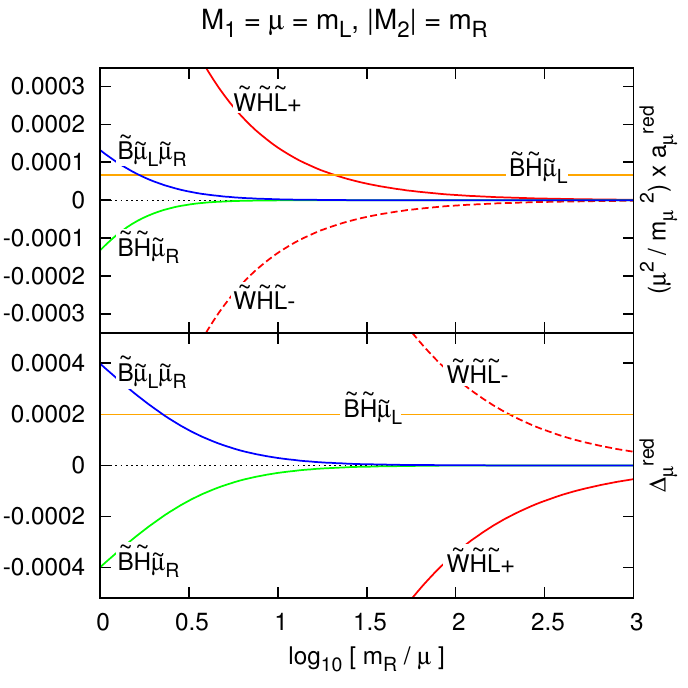}
\caption{}
\label{fig:linesBHL}
\end{subfigure}
\caption{
(\subref{fig:signBHL}) Signs of $\amuSUSY$ in the plane of $\MR/\mu$ versus $\Mone/\mu$,
for the two possible signs of $\Mtwo$. The colour coding is the same as in Fig.~\ref{fig:signs}.
(\subref{fig:linesBHL}) Individual behaviour of the contributions to $\amured$ and $\Deltamured$
as a function of the mass ratio $\MR/\mu$. The presentation is analogous to Fig.~\ref{fig:fivelines}.
}
\label{fig:BHL}
\end{figure}
Finally we investigate the positive $\amuSUSY$ region with $\BHL$ dominance,
which cannot be seen for the most part in the previous graphs. 
This dominance can be realized under the hierarchy condition $\Mone, \mu, \ML \ll |\Mtwo|, \MR$,
which is satisfied in the small white region in the bottom left area of Fig.~\ref{fig:signs}(\subref{fig:signs4}). 
To better understand the $\BHL$ dominance,
we change the plot axes to the ones of Fig.~\ref{fig:BHL}(\subref{fig:signBHL}).
Here the sign of $\amuSUSY$ as well as information about which contributions dominate
in $\amured$ and $\Deltamured$ are displayed in the plane of $m_R/\MUE$ versus $\Mone/\MUE$. 
Fig.~\ref{fig:BHL}(\subref{fig:linesBHL}) corresponds to the $y$-axis
of Fig.~\ref{fig:BHL}(\subref{fig:signBHL}) and shows the individual behaviour of each contribution. 
As we can see, the $\BHL$ contributions dominate only for extreme mass ratios $\MR/\mu\gsim100$.

\subsection{\boldmath Magnitude of \texorpdfstring{$\amuSUSY$}{amuSUSY} and constraints on Yukawa couplings}
\label{sec:magnitude}

Having understood the parameter regions which lead to positive $\amuSUSY$,
we now turn our interest to its magnitude and also take into account other constraints on the parameter space. 
As discussed in the Introduction, we can write 
\begin{align}
\amuSUSY &= C\frac{m_\mu^2}{M_{\text{SUSY,min}}^2},
&
C &= {\cal O}(1),
\label{DefC}
\end{align}
where $M_{\text{SUSY,min}}$ is the smallest among the five relevant
mass parameters $\Mone$, $|\Mtwo|$, $\mu$, and $m_{L,R}$ [and thus
similar but not equal to the lightest SUSY particle (LSP) mass since
the latter arises from diagonalization of mass matrices]. 
The coefficient $C$ is expected to be of order unity because of radiative muon mass generation. 
It depends only on the four independent mass ratios (up to terms suppressed by powers of $M_Z^2/M_{\text{SUSY}}^2$). 
Hence for each given set of mass ratios we can pose two questions:
\begin{enumerate}
\item What is the value of $C$?
\item What is the minimum SUSY mass for which
  $\amuSUSY=28.7\times10^{-10}$?
\end{enumerate}
\begin{figure}[t!]
\begin{subfigure}[b]{\textwidth}
\includegraphics[width=\textwidth]{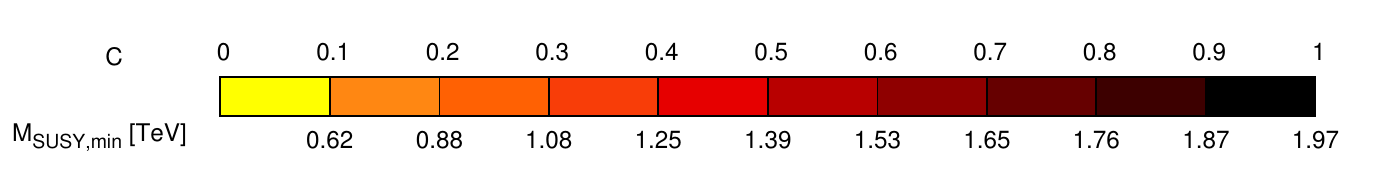}
\end{subfigure}
\begin{subfigure}[b]{0.5\textwidth}
\includegraphics[width=\textwidth]{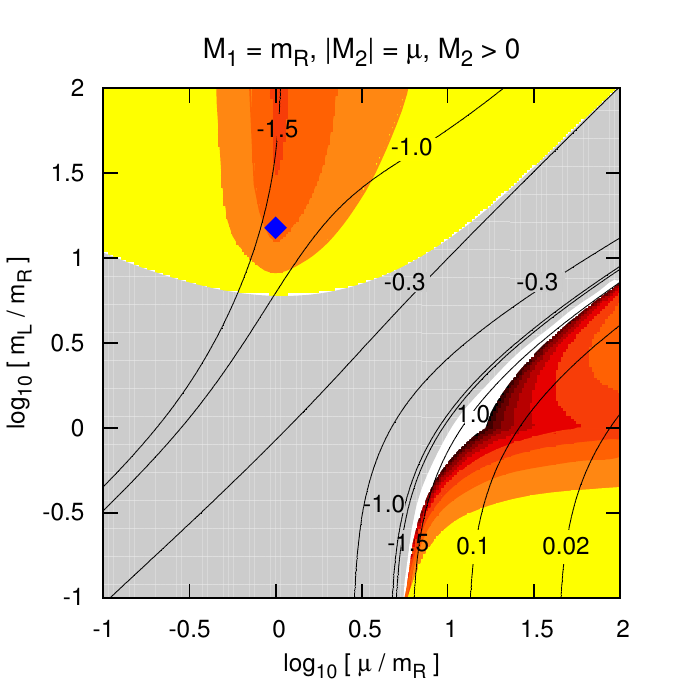}
\caption{}
\label{fig:gNYpos}
\end{subfigure}
\begin{subfigure}[b]{0.5\textwidth}
\includegraphics[width=\textwidth]{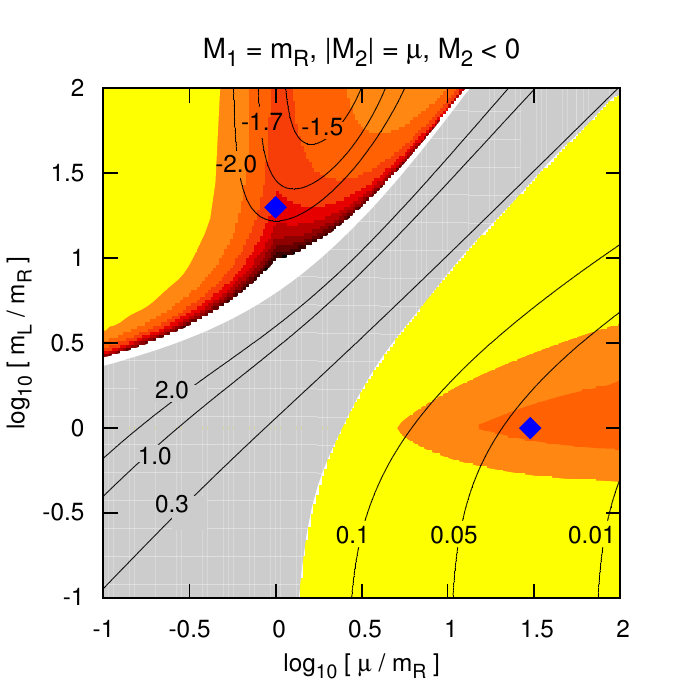}
\caption{}
\label{fig:gNYneg}
\end{subfigure}
\caption{
Magnitude of $\amuSUSY$ and muon Yukawa coupling $y_\mu$ in the plane of $\ML/\MR$ versus $\MUE/\MR$,
for positive $\Mtwo$ (left) and negative $\Mtwo$ (right). 
The colour coding corresponds to the different values of the coefficient $C$ defined in Eq.~\eqref{DefC} and the
equivalent minimum SUSY mass which leads to agreement with the current deviation via $\amuSUSY = 28.7\times10^{-10}$. 
The black contours correspond to the indicated values of the muon Yukawa coupling
resulting from Eq.~\eqref{eq:mmufromSE}. The grey regions are excluded by negative $\amuSUSY$. 
In the white regions the denominator $\Deltamured$ undergoes a sign change,
and the perturbation theory becomes untrustworthy due to large $y_\mu$.
Benchmark parameter points introduced in Table~\ref{tab:BMPoints} are marked with blue squares.}
\label{fig:gNY}
\end{figure}
The answers are given in Fig.~\ref{fig:gNY} for the special choices $\Mone=\MR$ and $|\Mtwo|=\mu$
which have already been used in Fig.~\ref{fig:signs}(\subref{fig:signs1}).
The figures also show the behaviour of the muon Yukawa coupling, 
which is proportional to $1/\Deltamured$, see Eq.~\eqref{eq:additionalformulas}.
The four most interesting regions are the  ``large~$\mu$-limit'' and
``$\tilde{\mu}_R$-dominance'' regions with either positive or negative
$\Mtwo$. These are located in the right end and middle top areas of the two
plots and correspond to the white regions of
Fig.~\ref{fig:signs}(\subref{fig:signs1}) with a positive
$\amuSUSY$. In detail Fig.~\ref{fig:gNY} shows the following.
\begin{itemize}
\item
The first important observation is that in the centre of the ``large~$\mu$-limit'' and
``$\tilde{\mu}_R$-dominance'' regions, $C$ lies  between $0.2$ and
$0.4$; the corresponding minimum SUSY mass is around $\unit{1}{\tera\electronvolt}$. 
$C$ does not quite reach unity because of the behaviour of the loop
functions in $\Deltamured$ and $\amured$.
\item
In both regions, the magnitude of $C$, the
minimum SUSY mass, and the muon Yukawa coupling depend on the sign of
$\Mtwo$. This dependence arises from interference of the dominant
contributions with subdominant wino contributions, which
can be either constructive in $\Deltamured$ and destructive in $\amured$ or vice versa.
\item
There are small white strips in parameter space with larger $C$ and
minimum SUSY masses.
In these strips $\Deltamured$ undergoes a sign change and the Yukawa
coupling becomes infinite.
Here perturbation theory is not trustworthy since two-loop
effects are non-negligible due to the accidental cancellation of the
one-loop contributions. We exclude these regions from further
discussion.
\item
Of particular interest are the regions with smaller Yukawa
couplings, i.e.\ the ``large $\mu$-limit'' region for negative
$\Mtwo$ and the ``$\tilde{\mu}_R$-dominance'' region for positive
$\Mtwo$. Here $C$ can still increase up to $0.3$, and the minimum
SUSY mass reaches up to $\unit{1.1}{\tera\electronvolt}$.
The opposite sign choices lead to the strips where $y_\mu$ diverges.
In the ``large $\mu$-limit'' the muon Yukawa coupling
is always proportional to $1/\mu$ and therefore its magnitude is comparatively
small and decreasing as $\mu$ increases. 
In the regions of ``$\tilde{\mu}_R$-dominance'' $|y_\mu|$ is generally much larger. 
Eq.~\eqref{eq:sigmamuapprox} allows to deduce that for  ``$\tilde{\mu}_R$-dominance'',
$|y_\mu|$ cannot become smaller than $0.78$ for negative $\Mtwo$ and $0.12$ for positive $\Mtwo$, respectively.
\end{itemize}

\begin{figure}[t!]
\begin{subfigure}[b]{\textwidth}
\centering
\includegraphics[width=0.743\textwidth]{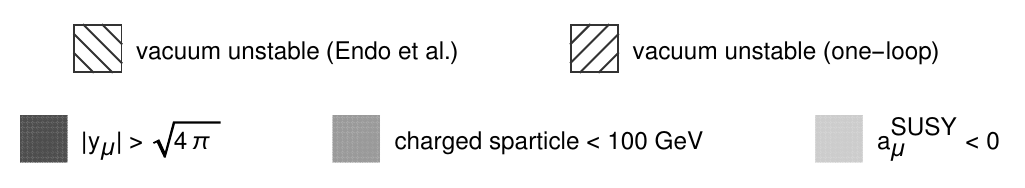}
\vspace{0.02\textwidth}
\end{subfigure}
\begin{subfigure}[b]{0.5\textwidth} 
\includegraphics[width=\textwidth]{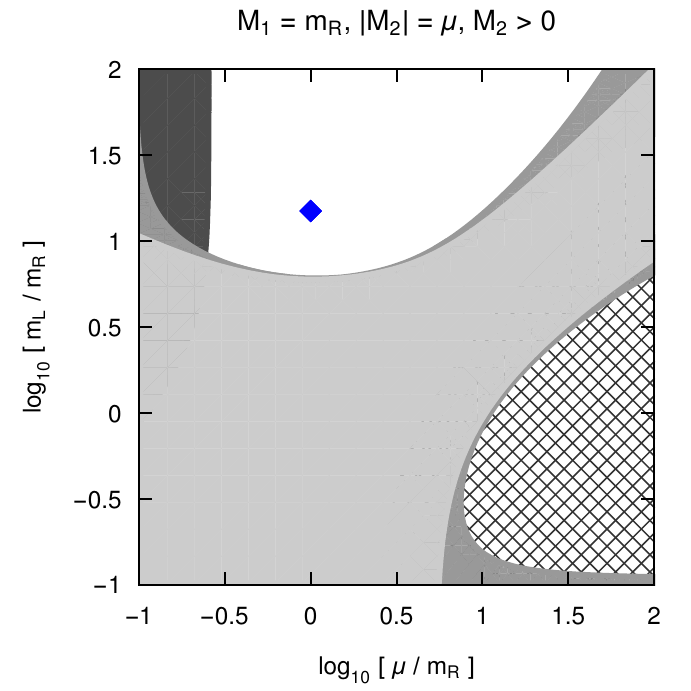}
\caption{}
\label{fig:exclplotpos}
\end{subfigure}
\begin{subfigure}[b]{0.5\textwidth}
\includegraphics[width=\textwidth]{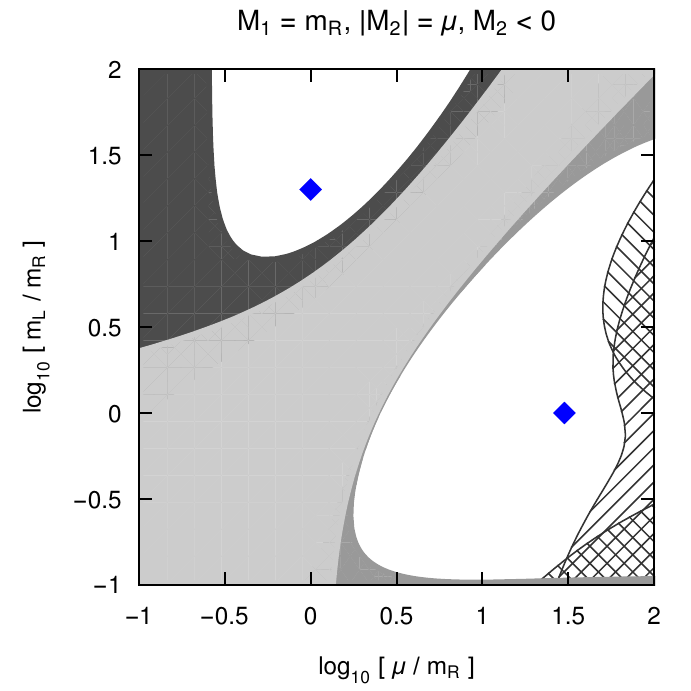}
\caption{}
\label{fig:exclplotneg}
\end{subfigure}
\caption{Excluded parameter regions in the plane of $\ML/\MR$ versus $\MUE/\MR$,
for positive $\Mtwo$ (left) and negative $\Mtwo$ (right).
The areas corresponding to the different constraints are plotted in the same order as they appear in the legend.
Detailed explanations for each constraint can be found in the text.
Benchmark parameter points introduced in Table~\ref{tab:BMPoints} are marked with blue squares.}
\label{fig:Yukawaconstraints}
\end{figure}
Now, we consider constraints on the relevant parameter space for large $\amuSUSY$.
All previous plots are based on the mass-insertion approximation and depend only on mass ratios.
For each given set of mass ratios with positive $\amuSUSY$ there is a value
of the overall mass scale $M_{\text{SUSY}}$ for which $\amuSUSY$ agrees with experiment.
In the following discussions we always fix $M_{\text{SUSY}}$ to that particular scale.

Fig.~\ref{fig:Yukawaconstraints} shows the same parameter space as
Fig.~\ref{fig:gNY}, but displays regions in which charged SUSY masses become too small (medium grey),
the muon Yukawa coupling is non-perturbative (dark grey) or the electroweak vacuum becomes unstable (hatched).
In detail, the constraints are the following:
\begin{itemize}
\item The light grey regions, which have already been shown in Fig.~\ref{fig:gNY},
yield negative values of $\amuSUSY$ and are thus not of interest
with regard to an explanation of the discrepancy~\eqref{eq:amuexp-amusm}.
\item There are strong but model-dependent collider bounds on the
  masses of charged SUSY particles. Here we are interested in very
  large electroweak SUSY masses, which are far above the LHC
  limits. Therefore we show as the medium grey regions in
  Fig.~\ref{fig:Yukawaconstraints} an exemplary contour corresponding
  to a chargino or smuon which is lighter than $\unit{100}{\giga\electronvolt}$.
\item In the dark grey regions  the muon Yukawa coupling violates
  perturbativity, $|y_\mu|>\sqrt{4\pi}$. Generally,
we regard the model as a pure low-energy model and do not require the
stronger constraint that there be no Landau poles at higher scales;
such an analysis has been done in Refs.~\cite{Dobrescu:2010mk,Altmannshofer:2010zt}. 
It should be further noted that the observable couplings of the lightest Higgs boson are SM-like
up to corrections suppressed by powers of the SUSY scale. 
This is true independently of the value of the fundamental Yukawa coupling due to decoupling.
\item As discussed in Refs.~\cite{Endo:2013lva,Hisano:2010re,Kitahara:2012pb,Carena:2012mw,Kitahara:2013lfa},
the combination $y_\ell \mu$ is limited by the requirement that the electroweak vacuum is metastable
with a lifetime longer than the age of the universe. 
This constrains in particular the parameter regions with large $\mu$ as already stressed in Ref.~\cite{Endo:2013lva}. 
The regions which are excluded by the fitting formula from this reference applied to the muon sector are shown in Fig.~\ref{fig:Yukawaconstraints}.
Since it is possible that $y_\ell \gsim y_t$ in our scenario,
the $\mathcal{O}(y_\ell^n)$ counterpart of the standard $\mathcal{O}(y_t^n)$
correction to the effective potential might be significant.
To scrutinize the effect of the former,
we add to the effective potential the one-loop contributions
(see e.g.\ Eq.~(1.2) of Ref.~\cite{Martin:2002iu})
in the gaugeless limit, with the renormalization scale $Q = \sqrt{m_L m_R}$.
They include the (s)lepton- and the Higgs(ino)- as well as the (s)top-loop contributions.
We then evaluate the Euclidean action numerically using the method
presented in Ref.~\cite{jaesmethod}.
In principle, the metastability bound on $y_\ell \mu$ depends on
the pseudo-scalar Higgs mass $M_A$ affecting the one-loop correction to
the slepton squared mass.
Motivated by the $B$-physics constraints,
we set $M_A=\unit{50}{\tera\electronvolt}$,
a high enough value for extra Higgs states to decouple.
By applying the specified procedure to the muon sector,
we find that the regions shown in Fig.~\ref{fig:Yukawaconstraints} are excluded because of
an unstable electroweak vacuum.
\end{itemize}
According to the results shown in Fig.~\ref{fig:Yukawaconstraints},
the requirement of vacuum metastability poses strong constraints on
parameter space, in particular on the ``large $\mu$-limit'' region. This
region is completely excluded for positive $\Mtwo$, while for negative
$\Mtwo$ the ratio $\mu/\MR$ must not be too large. The
``$\tilde{\mu}_R$-dominance'' region is not constrained by vacuum metastability.
The two different approaches to the vacuum metastability yield similar results
but the influence of the additional terms is visible in Fig.~\ref{fig:Yukawaconstraints}(\subref{fig:exclplotneg}).

The Yukawa coupling of the $\tau$ lepton, which is another quantity of interest,
can be determined similarly to $y_\mu$. There is a strong correlation if the stau masses are assumed
to be equal to the smuon masses; without this assumption the two Yukawa couplings are basically independent,
and vacuum metastability from the tau sector does not further constrain the five-dimensional parameter space.
However, a detailed analysis should take into account higher orders in the large tau Yukawa coupling
and is beyond the scope of this paper.

Thus we find that out of the four interesting regions in Fig.~\ref{fig:gNY}, three survive.
Now we briefly comment on the comparison between the approximate results for $y_\mu$
as well as $\amuSUSY$ obtained from Eqs.~(\ref{eq:sigmamuparts},\,\ref{eq:amuparts})
and their exact results according to the Appendix.
In the three viable regions described above, the deviation amounts to at most a few percent.
Outside these regions the approximation becomes worse for very small SUSY masses
and in the small strips with divergent muon Yukawa coupling.

\section{Conclusions}
\label{sec:con}

\newcommand{\tabtev}{}
\begin{table}
\centerline{
\begin{tabular}{ccccc|cc|c|c}
 \hline
 $\mu\tabtev$ & $\Mone\tabtev$ & $\Mtwo\tabtev$ &
 $\ML\tabtev$ & $\MR\tabtev$ & $\amuSUSY/10^{-9}$ & $\phantom{-}y_\mu$ &
 Figure & Characteristic\\ \hline
 $30$ & $1$ & $-30\phantom{-}$ & $1$ & $1$ & $2.80$ &
 $\phantom{-}0.04$ &\ref{fig:gNYneg}&large $\mu$-limit\\
 $15$ & $1$ & $-1\phantom{-}$ & $1$ & $1$ & $3.01$ & $\phantom{-}0.09$ &--&large $\mu$-limit\\
 $1$ & $1$ & $\phantom{-}1\phantom{-}$ & $15$ & $1$ & $2.64$ & $-1.37$
 &\ref{fig:gNYpos}&$\tilde{\mu}_R$-dominance\\
 $1$ & $1$ & $\phantom{-}30\phantom{-}$ & $30$ & $1$ & $2.77$ &
 $-1.18$ &--&$\tilde{\mu}_R$-dominance\\
 $1.3$ & $1.3$ & $-1.3\phantom{-}$ & $26$ & $1.3$ & $2.90$ & $-1.89$
 &\ref{fig:gNYneg}&$\tilde{\mu}_R$-dominance\\
 \hline
\end{tabular}
}
\caption{\label{tab:BMPoints}Benchmark parameter points. All masses
  are given in $\tera\electronvolt$.}
\end{table}

We have studied the muon magnetic moment in the MSSM for
$\tan\beta\to\infty$. This is a viable limit in which all down-type
masses are generated by loop-induced couplings to the ``wrong'' Higgs $H_u$.
The scenario can also be viewed independently of supersymmetry,
as a simplified model which realizes the general idea of radiative muon mass generation.

The SUSY contribution to the muon anomalous magnetic moment is given by Eq.~\eqref{amuSUSYratio} as the
ratio of two one-loop quantities. 
For this reason the scenario leads to large
$a_\mu$ with TeV-scale new physics masses, and it behaves qualitatively very differently from the standard
MSSM with $\tan\beta\lsim50$. E.g.\ the signs of gaugino and Higgsino
masses drop out between numerator and denominator and cannot be used
to adjust the overall sign of $\amuSUSY$. For equal masses, $\amuSUSY$
is strictly negative, see Eq.~\eqref{negativeamuSUSY}. In general,
the sign of $\amuSUSY$ depends on the ratios 
between the five relevant mass parameters;
Figs.~\ref{fig:fivelines}--\ref{fig:BHL} give a comprehensive
explanation of which mass ratios lead to positive $\amuSUSY$. 

The three most promising parameter regions with positive $\amuSUSY$
are the ``large~$\mu$-limit'' for negative $\Mtwo$ and the
``$\tilde{\mu}_R$-dominance'' region for positive or negative
$\Mtwo$. The ``large $\mu$-limit'' region with positive $M_2$ is excluded
because of vacuum instability. In all three viable cases,
large contributions to $\amuSUSY$ are possible without
conflicts with other experimental constraints, see
Figs.~\ref{fig:gNY}, \ref{fig:Yukawaconstraints}. The discrepancy~\eqref{eq:amuexp-amusm}
can be fully explained with a lightest SUSY mass 
above $\unit{1}{\tera\electronvolt}$. If this scenario is correct, the
LHC would not be able to find the SUSY particles relevant for
$a_\mu$. However,  Sec.~\ref{sec:expconstraints} shows that
the scenario could potentially influence and thus be tested in a
multitude of observables in $B$-physics and Higgs physics.

To conclude, we summarize the behaviour in three simple ways.
Table~\ref{tab:BMPoints} lists several benchmark parameter points with SUSY mass parameters
at $\unit{1}{\tera\electronvolt}$ or higher and gives
the resulting $\amuSUSY$ and muon Yukawa coupling. Two points correspond to the
``large $\mu$-limit'' with negative $\Mtwo$ and three to
the ``$\tilde{\mu}_R$-dominance'' region with positive or negative
$\Mtwo$. Some points involve only one
heavy and four lighter SUSY masses; for these large $\amuSUSY$ is possible for less extreme mass
splitting than for the ones with two heavy and three lighter masses. The benchmark points
further illustrate that the Yukawa 
coupling is much larger in the case of ``$\tilde{\mu}_R$-dominance''.

\begin{figure}[t]
\begin{subfigure}[b]{0.5\textwidth} 
\includegraphics[scale=1.0]{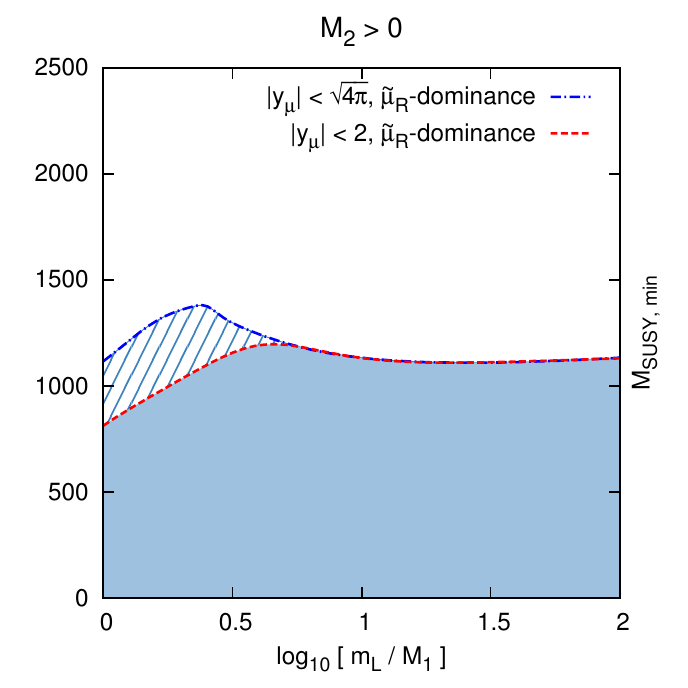}
\caption{}
\end{subfigure}
\begin{subfigure}[b]{0.5\textwidth}
\includegraphics[scale=1.0]{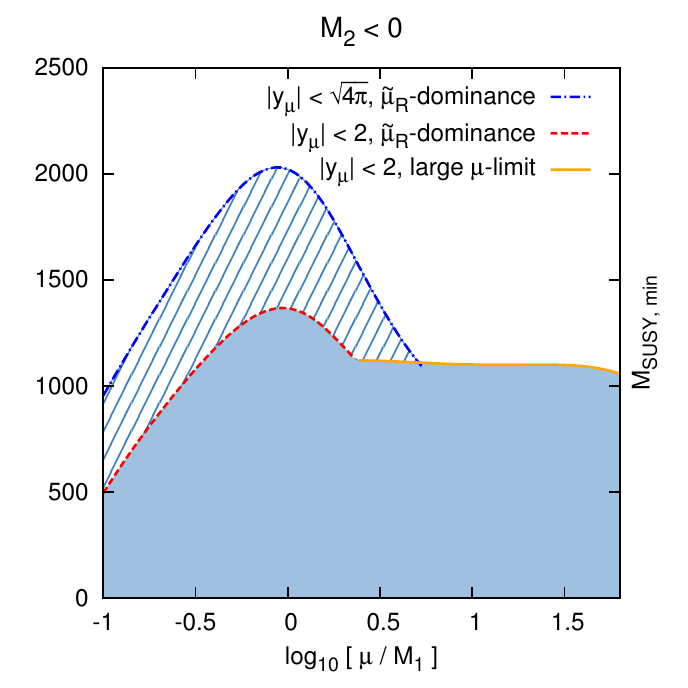}
\caption{}
\end{subfigure}
\caption{Results of scans over all
  five relevant SUSY mass parameters. The plots show the maximum
  values of $M_{\text{SUSY,min}}$ for which the deviation~\eqref{eq:amuexp-amusm}
  can be explained, also indicating the
  characteristic of the parameter point for which the maximum is
  achieved. In some regions the achievable maximum depends on the
  cutoff on the magnitude of the muon Yukawa coupling, as indicated; in the
  other regions the maximum is achieved for smaller Yukawa couplings.
  The scans take into account the constraint of vacuum 
  metastability (one-loop), as discussed in Sec.~\ref{sec:magnitude}, and in the right plot
  all investigated parameter points with $\log_{10}(\mu/\Mone)>1.8$ are excluded by this.
}
\label{fig:Scanplots}
\end{figure}

Fig.~\ref{fig:Scanplots} shows the results of two scans of the
five-dimensional parameter space: plotted are the largest possible
values of the minimum SUSY mass $M_\text{SUSY,min}$ which lead to
agreement with Eq.~\eqref{eq:amuexp-amusm}. The left panel shows the
results as a function of $\log_{10}(m_L/\Mone)$ (for positive $\Mtwo$),
it corresponds to the $y$-axis of Fig.~\ref{fig:gNY}(\subref{fig:gNYpos}); 
the right panel as a function of  $\log_{10}(\mu/\Mone)$ (for negative $\Mtwo$),
it corresponds to the $x$-axis of Fig.~\ref{fig:gNY}(\subref{fig:gNYneg}).
The results correspond to the regions presented in Fig.~\ref{fig:gNY} and
are just slightly higher than the values obtained there with fixed mass ratios.

Finally, for the regions of interest, a simple approximation in the style
of Eq.~\eqref{negativeamuSUSY} can be derived: for the ``large $\mu$-limit'' region,
we consider the limit $|\mu|\gg |\Mone|=m_L=m_R\equiv M_{\text{SUSY}}$,
and for the ``$\tilde{\mu}_R$-dominance'' region we consider the limit
$m_L\gg |\mu|=|\Mone|=m_R\equiv M_{\text{SUSY}}$.
In both of these limits, we obtain
\begin{align}
\amuSUSY\approx37\times10^{-10} \,
 \left(\frac{\unit{1}{\tera\electronvolt}}{M_{\text{SUSY}}}\right)^2.
\end{align}

\subsection*{Acknowledgements}

J.P. acknowledges support from the MEC and FEDER (EC) Grants
FPA2011--23596 and the Generalitat Valenciana under grant PROMETEOII/2013/017.
We also acknowledge support from DFG Grants
STO/876/2-1, STO/876/6-1.

\begin{appendix}
\section{Appendix}

In this Appendix we provide the full one-loop results for the muon
self energy and the magnetic moment, defined in
Eqs.~(\ref{eq:mmufromSE},\,\ref{eq:amured}). We use the notation of
Ref.~\cite{Fargnoli:2013zia}, which is similar to the ones from
Refs.~\cite{MartinWells,Stockinger:2006zn}. Note that in the standard
case of the MSSM with moderate $\tan\beta\lsim50$, 
the combination $m_\mu\tan\beta$ frequently appears. An important
difference is that here we have to take into account the distinction
between the tree-level and the physical muon mass, $m_\mu^\text{tree}\ne
m_\mu$, and we have to use the identification~\eqref{eq:tbeff},
\begin{align}
m_\mu^{\text{tree}}\tan\beta=m_\mu\tbeff
= y_\mu v_u =
  y_\mu \frac{\sqrt{2} M_W}{g_2},
  \label{eq:mmu0tanbeta}
\end{align}
which is exactly valid. Here $g_2$ denotes the SU(2) gauge coupling.

The relevant interaction terms for chargino--sneutrino and
neutralino--smuon interactions with muons can be written as
\begin{align}
  \mathcal{L}_\text{int}=
  \overline{\tilde\chi_i^-}
  \left(c_{i\tilde\nu_\mu}^L P_L + c_{i\tilde\nu_\mu}^R P_R\right)
  \mu \, \tilde\nu_\mu^\dagger
  +\overline{\tilde\chi_j^0}
  \left(n_{j\tilde\mu_k}^L P_L + n_{j\tilde\mu_k}^R P_R\right)
  \mu \, \tilde\mu_k^\dagger + \text{h.\,c.}\,,
\end{align}
where the coupling coefficients are defined as
\begin{subequations}
\label{eq:cLRnLR}
\begin{align}
  c_{i\tilde\nu_\mu}^L&=-g_2 V_{i1}^\ast,\\
  c_{i\tilde\nu_\mu}^R&=y_\mu U_{i2},\\
  n_{j\tilde\mu_k}^L&=\frac{1}{\sqrt{2}}
    \left(g_1 N_{j1}^\ast+g_2 N_{j2}^\ast\right)U_{k1}^{\tilde\mu}-
    y_\mu N_{j3}^\ast U_{k2}^{\tilde\mu},\\
  n_{j\tilde\mu_k}^R&=-\sqrt{2} g_1 N_{j1} U_{k2}^{\tilde\mu}
    -y_\mu N_{j3} U_{k1}^{\tilde\mu}.
\end{align}
\end{subequations}
The unitary matrices $U,V,N$ diagonalize the chargino and neutralino
mass matrices; $U^{\tilde\mu}$ is used to diagonalize the smuon mass matrix.
The resulting smuon mass eigenvalues are given by
\begin{align}
  m_{\tilde\mu_{1/2}}^2=
  \frac{1}{2}\Bigg(m_L^2+m_R^2+\frac{M_Z^2}{2}
  \pm\sqrt{\left[m_L^2-m_R^2+M_Z^2 \left(\frac{1}{2}-2 s_W^2
  \right)\right]^2+8 y_\mu^2 |\mu|^2 \frac{M_W^2}{g_2^2}}\Bigg).
  \label{eq:msmu}
\end{align}
A term containing the trilinear coupling vanishes because of $v_d=0$,
and in the last term we have used Eq.~\eqref{eq:mmu0tanbeta}.
The occurring SUSY parameters are the
higgsino mass parameter $\mu$, the gaugino masses $M_1$ as well as $M_2$ and
the soft mass parameters for the second-generation slepton doublet and singlet
$m_L$ and $m_R$, respectively.

With this notation, the MSSM one-loop contributions to the on-shell
muon self-energy from chargino--sneutrino or 
neutralino--smuon loops can be expressed as
$\Sigma_\mu^\text{MSSM} =
\Sigma_\mu^{\tilde\chi^\pm} + \Sigma_\mu^{\tilde\chi^0}=-y_\mu v_u\Deltamured$, with
\begin{subequations}
\label{eq:sigmamuchipm0}
\begin{align}
  \Sigma_\mu^{\tilde\chi^\pm}&=
  \frac{1}{16\pi^2}\sum_{i} m_{\tilde\chi_i^-}
  \operatorname{Re}\left[c_{i\tilde\nu_\mu}^L(c_{i\tilde\nu_\mu}^R)^\ast\right]
  B_0(m_{\tilde\chi_i^-},m_{\tilde\nu_\mu}),\\
  \Sigma_\mu^{\tilde\chi^0}&=
  \frac{1}{16\pi^2}\sum_{j,k} m_{\tilde\chi_j^0}
  \operatorname{Re}\left[n_{j\tilde\mu_k}^L(n_{j\tilde\mu_k}^R)^\ast\right]
  B_0(m_{\tilde\chi_j^0},m_{\tilde\mu_k}),
\end{align}
\end{subequations}
where terms suppressed by $m_\mu/M_\text{SUSY}$ or by powers of
$\tan\beta$ are neglected. In dimensional
regularization with parameter $\epsilon=(4-D)/2$ and regularization scale
$\mu_R$, the one-loop integral $B_0$ is given by
\begin{align}
  B_0(a,b)=\Delta+1+\frac{1}{b^2-a^2}
  \left(a^2 \ln\frac{a^2}{\mu_R^2} - b^2 \ln\frac{b^2}{\mu_R^2}\right)
  +\mathcal{O}(\epsilon),
\end{align}
with the common quantity
\begin{align}
  \Delta=\frac{1}{\epsilon}+\ln 4\pi-\gamma_E,
\end{align}
containing the Euler-Mascheroni constant $\gamma_E$. Owing to the
unitarity of the mixing matrices, in Eqs.~\eqref{eq:sigmamuchipm0} the
UV divergences as well as the dependence on $\mu_R$ vanish.

Because of $v_d=0$, the tree-level mass of the muon (as well as the other
charged leptons and down-type quarks) equals zero. Therefore, the entire muon
mass has to be generated at the loop level, which means
\begin{align}
  m_\mu = -\Sigma_\mu^\text{MSSM}.
\end{align}
For a given parameter point, this equation can be used to determine the muon
Yukawa coupling $y_\mu$. Note that the self energy is approximately
linear in $y_\mu$, as suggested by the definition of $\Deltamured$ in
Eq.~\eqref{eq:mmufromSE}. But the Yukawa coupling also enters nonlinearly,
through quadratic terms in
$n_{j\tilde\mu_k}^L(n_{j\tilde\mu_k}^R)^\ast$, and via the smuon
masses~\eqref{eq:msmu}. Therefore, it has to be
determined numerically in an exact calculation. As shown in
Sec.~\ref{sec:model}, however, the nonlinear terms vanish in the
mass-insertion approximation. 

The SUSY one-loop diagrams contributing to the anomalous magnetic moment of the
muon are the same as for the muon self-energy, except for the fact that now a
photon couples to the charged SUSY particle in the loop, i.e.\ the chargino or
the smuon. Again neglecting terms suppressed by $\tan\beta$, the result reads
$\amuSUSY = a_\mu^{\tilde\chi^\pm} + a_\mu^{\tilde\chi^0}=({y_\mu v_u}/{m_\mu})
\amured$, with
\begin{subequations}
\label{eq:amuchipm0}
\begin{align}
  a_\mu^{\tilde\chi^\pm}&= \phantom{-}
  \frac{m_\mu}{24\pi^2} \sum_{i} \frac{m_{\tilde\chi_i^-}}{m_{\tilde\nu_\mu}^2}
  \operatorname{Re}\left[c_{i\tilde\nu_\mu}^L (c_{i\tilde\nu_\mu}^R)^\ast\right]
  F_2^C(x_i),\\
  a_\mu^{\tilde\chi^0}&=
  -\frac{m_\mu}{48\pi^2} \sum_{j,k} \frac{m_{\tilde\chi_j^0}}{m_{\tilde\mu_k}^2}
  \operatorname{Re}\left[n_{j\tilde\mu_k}^L (n_{j\tilde\mu_k}^R)^\ast\right]
  F_2^N(x_{jk}),
\end{align}
\end{subequations}
where the mass ratios are given by $x_i = m_{\tilde\chi_i^-}^2 /
m_{\tilde\nu_\mu}^2$ and $x_{jk} = m_{\tilde\chi_j^0}^2 / m_{\tilde\mu_k}^2$.
The dimensionless loop functions have been introduced in
Ref.~\cite{MartinWells} as
\begin{subequations}
\begin{align}
  F_2^C(x)&=\frac{3}{2(1-x)^3}\left[-3+4x-x^2-2\ln x\right],\\
  F_2^N(x)&=\frac{3}{(1-x)^3}\left[1-x^2+2x\ln x\right],
\end{align}
\end{subequations}
normalized such that $F_2^C(1)=F_2^N(1)=1$. The formulas look
identical to the respective terms in the standard case with moderate
$\tan\beta$, but like for the self energy, there are nonlinear terms
in $y_\mu$, which usually are neglected because of the
$m_\mu/M_\text{SUSY}$ suppression. In our case, these terms are only
suppressed as $M_{W,Z}/M_\text{SUSY}$; hence they are much more
significant in an exact calculation. Nevertheless, like for the self
energy, in the mass-insertion approximation, $\amuSUSY$ is linear in
the Yukawa coupling.

\end{appendix}


\begin{thebibliography}{AA}
\bibitem{Bennett:2006fi}
  G.~W.~Bennett {\it et al.}  [Muon g-2 Collaboration],
  Phys.\ Rev.\ D {\bf 73} (2006) 072003
  [hep-ex/0602035].
\bibitem{Carey:2009zzb}
  R.~M.~Carey, K.~R.~Lynch, J.~P.~Miller, B.~L.~Roberts, W.~M.~Morse, Y.~K.~Semertzides, V.~P.~Druzhinin and B.~I.~Khazin {\it et al.},
  FERMILAB-PROPOSAL-0989;
  B.~L.~Roberts,
  Chin.\ Phys.\ C {\bf 34} (2010) 741
  [arXiv:1001.2898 [hep-ex]];
  J.~Grange {\it et al.}  [Muon g-2 Collaboration],
  arXiv:1501.06858 [physics.ins-det].
\bibitem{Iinuma:2011zz}
  H.~Iinuma [J-PARC New g-2/EDM experiment Collaboration],
  J.\ Phys.\ Conf.\ Ser.\  {\bf 295} (2011) 012032.
\bibitem{JegerlehnerNyffeler}
  F.~Jegerlehner and A.~Nyffeler,
  Phys.\ Rept.\  {\bf 477} (2009) 1
  [arXiv:0902.3360 [hep-ph]].
\bibitem{Miller:2012opa}
  J.~P.~Miller, E.~de Rafael, B.~L.~Roberts and D.~St\"ockinger,
  Ann.\ Rev.\ Nucl.\ Part.\ Sci.\  {\bf 62} (2012) 237.
\bibitem{Kinoshita2012}
  T.~Aoyama, M.~Hayakawa, T.~Kinoshita and M.~Nio,
  Phys.\ Rev.\ Lett.\  {\bf 109} (2012) 111808
  [arXiv:1205.5370 [hep-ph]].
\bibitem{Kataev:2012kn}
  A.~L.~Kataev,
  Phys.\ Rev.\ D {\bf 86} (2012) 013010
  [arXiv:1205.6191 [hep-ph]].
\bibitem{SteinhauserQED}
  A.~Kurz, T.~Liu, P.~Marquard and M.~Steinhauser,
  Nucl.\ Phys.\ B {\bf 879} (2014) 1
  [arXiv:1311.2471 [hep-ph]];
  R.~Lee, P.~Marquard, A.~V.~Smirnov, V.~A.~Smirnov and M.~Steinhauser,
  JHEP {\bf 1303} (2013) 162
  [arXiv:1301.6481 [hep-ph]].
\bibitem{Gnendiger:2013pva}
  C.~Gnendiger, D.~St{\"o}ckinger and H.~St{\"o}ckinger-Kim,
  Phys.\ Rev.\ D {\bf 88} (2013) 053005
  [arXiv:1306.5546 [hep-ph]].
\bibitem{Davier}
  M.~Davier, A.~Hoecker, B.~Malaescu and Z.~Zhang,
  Eur.\ Phys.\ J.\ C {\bf 71} (2011) 1515
   [Erratum-ibid.\ C {\bf 72} (2012) 1874]
  [arXiv:1010.4180 [hep-ph]].
\bibitem{HMNT}
  K.~Hagiwara, R.~Liao, A.~D.~Martin, D.~Nomura and T.~Teubner,
  J.\ Phys.\ G {\bf 38} (2011) 085003
  [arXiv:1105.3149 [hep-ph]].
\bibitem{Benayoun:2012wc}
  M.~Benayoun, P.~David, L.~DelBuono and F.~Jegerlehner,
  Eur.\ Phys.\ J.\ C {\bf 73} (2013) 2453
  [arXiv:1210.7184 [hep-ph]].
\bibitem{JegerlehnerSzafron}
  F.~Jegerlehner and R.~Szafron,
  Eur.\ Phys.\ J.\ C {\bf 71} (2011) 1632
  [arXiv:1101.2872 [hep-ph]].
\bibitem{Kurz:2014wya}
  A.~Kurz, T.~Liu, P.~Marquard and M.~Steinhauser,
  Phys.\ Lett.\ B {\bf 734} (2014) 144
  [arXiv:1403.6400 [hep-ph]].
\bibitem{Blum:2013xva}
  T.~Blum, A.~Denig, I.~Logashenko, E.~de Rafael, B.~L.~Roberts, T.~Teubner and G.~Venanzoni,
  arXiv:1311.2198 [hep-ph].
\bibitem{Benayoun:2014tra}
  M.~Benayoun, J.~Bijnens, T.~Blum, I.~Caprini, G.~Colangelo, H.~Czyz, A.~Denig and C.~A.~Dominguez {\it et al.},
  arXiv:1407.4021 [hep-ph].
\bibitem{Colangelo:2014pva}
  G.~Colangelo, M.~Hoferichter, B.~Kubis, M.~Procura and P.~Stoffer,
  Phys.\ Lett.\ B {\bf 738} (2014) 6
  [arXiv:1408.2517 [hep-ph]].
\bibitem{Masjuan:2014rea}
  P.~Masjuan,
  arXiv:1411.6397 [hep-ph].
\bibitem{CzM}
  A.~Czarnecki and W.~J.~Marciano,
  Phys.\ Rev.\ D {\bf 64} (2001) 013014
  [hep-ph/0102122].
\bibitem{Borzumati:1999sp}
  F.~Borzumati, G.~R.~Farrar, N.~Polonsky and S.~D.~Thomas,
  Nucl.\ Phys.\ B {\bf 555} (1999) 53
  [hep-ph/9902443].
\bibitem{Crivellin:2010ty}
  A.~Crivellin, J.~Girrbach and U.~Nierste,
  Phys.\ Rev.\ D {\bf 83} (2011) 055009
  [arXiv:1010.4485 [hep-ph]].
\bibitem{Thalapillil:2014kya}
  A.~Thalapillil and S.~Thomas,
  arXiv:1411.7362 [hep-ph].
\bibitem{tanbe} 
  M.~Carena, M.~Olechowski, S.~Pokorski and C.~E.~M.~Wagner,
  Nucl.\ Phys.\ {\bf B426} (1994) 269;
  L.~J.~Hall, R.~Rattazzi and U.~Sarid,
  Phys.\ Rev.\ {\bf D50} (1994) 7048;
  T.~Blazek, S.~Raby and S.~Pokorski,
  Phys.\ Rev.\  D {\bf 52} (1995) 4151;
  C.~Hamzaoui, M.~Pospelov and M.~Toharia,
  Phys.\ Rev.\  D {\bf 59} (1999) 095005;
  M.~S.~Carena, D.~Garcia, U.~Nierste and C.~E.~M.~Wagner,
  Nucl.\ Phys.\  B {\bf 577} (2000) 88;
  K.~S.~Babu and C.~F.~Kolda,
  Phys.\ Rev.\ Lett.\  {\bf 84} (2000) 228;
  G.~Isidori and A.~Retico,
  JHEP {\bf 0111} (2001) 001;
  A.~J.~Buras, P.~H.~Chankowski, J.~Rosiek and L.~Slawianowska,
  Nucl.\ Phys.\  B {\bf 659} (2003) 3;
  A.~Crivellin and U.~Nierste,
  Phys.\ Rev.\ D {\bf 81} (2010) 095007;
  A.~Crivellin, L.~Hofer and J.~Rosiek,
  JHEP {\bf 1107} (2011) 017.
\bibitem{Marchetti:2008hw}
  S.~Marchetti, S.~Mertens, U.~Nierste and D.~St\"ockinger,
  Phys.\ Rev.\ D {\bf 79} (2009) 013010
  [arXiv:0808.1530 [hep-ph]].
\bibitem{Dobrescu:2010mk}
  B.~A.~Dobrescu and P.~J.~Fox,
  Eur.\ Phys.\ J.\ C {\bf 70} (2010) 263
  [arXiv:1001.3147 [hep-ph]].
\bibitem{Altmannshofer:2010zt}
  W.~Altmannshofer and D.~M.~Straub,
  JHEP {\bf 1009} (2010) 078
  [arXiv:1004.1993 [hep-ph]].
\bibitem{Freitas:2014pua}
  A.~Freitas, J.~Lykken, S.~Kell and S.~Westhoff,
  JHEP {\bf 1405} (2014) 145
   [Erratum-ibid.\  {\bf 1409} (2014) 155]
  [arXiv:1402.7065 [hep-ph]].
\bibitem{MartinWells}
  S.~P.~Martin and J.~D.~Wells,
  Phys.\ Rev.\ D {\bf 64} (2001) 035003
  [hep-ph/0103067].
\bibitem{Stockinger:2006zn}
  D.~St\"ockinger,
  J.\ Phys.\ G {\bf 34} (2007) R45
  [hep-ph/0609168].
\bibitem{Cho:2011rk}
  G.-C.~Cho, K.~Hagiwara, Y.~Matsumoto and D.~Nomura,
  JHEP {\bf 1111} (2011) 068
  [arXiv:1104.1769 [hep-ph]].
\bibitem{Endo:2013bba}
  M.~Endo, K.~Hamaguchi, S.~Iwamoto and T.~Yoshinaga,
  JHEP {\bf 1401} (2014) 123
  [arXiv:1303.4256 [hep-ph]].
\bibitem{Endo:2013lva}
  M.~Endo, K.~Hamaguchi, T.~Kitahara and T.~Yoshinaga,
  JHEP {\bf 1311} (2013) 013
  [arXiv:1309.3065 [hep-ph]].
\bibitem{Endo:2013xka}
  M.~Endo, K.~Hamaguchi, S.~Iwamoto, T.~Kitahara and T.~Moroi,
  Phys.\ Lett.\ B {\bf 728} (2014) 274
  [arXiv:1310.4496 [hep-ph]].
\bibitem{Fargnoli:2013zda}
  H.~G.~Fargnoli, C.~Gnendiger, S.~Pa\ss{}ehr, D.~St\"ockinger and H.~St\"ockinger-Kim,
  Phys.\ Lett.\ B {\bf 726} (2013) 717
  [arXiv:1309.0980 [hep-ph]].
\bibitem{Fargnoli:2013zia}
  H.~G.~Fargnoli, C.~Gnendiger, S.~Pa\ss{}ehr, D.~St\"ockinger and H.~St\"ockinger-Kim,
  JHEP {\bf 1402} (2014) 070
  [arXiv:1311.1775 [hep-ph]].
\bibitem{Evans:2012hg}
  J.~L.~Evans, M.~Ibe, S.~Shirai and T.~T.~Yanagida,
  Phys.\ Rev.\ D {\bf 85} (2012) 095004
  [arXiv:1201.2611 [hep-ph]].
\bibitem{Ibe:2012qu}
  M.~Ibe, S.~Matsumoto, T.~T.~Yanagida and N.~Yokozaki,
  JHEP {\bf 1303} (2013) 078
  [arXiv:1210.3122 [hep-ph]].
\bibitem{Ibe:2013oha}
  M.~Ibe, T.~T.~Yanagida and N.~Yokozaki,
  JHEP {\bf 1308} (2013) 067
  [arXiv:1303.6995 [hep-ph]].
\bibitem{Bhattacharyya:2013xma}
  G.~Bhattacharyya, B.~Bhattacherjee, T.~T.~Yanagida and N.~Yokozaki,
  Phys.\ Lett.\ B {\bf 730} (2014) 231
  [arXiv:1311.1906 [hep-ph]].
\bibitem{Mohanty:2013soa}
  S.~Mohanty, S.~Rao and D.~P.~Roy,
  JHEP {\bf 1309} (2013) 027
  [arXiv:1303.5830 [hep-ph]].
\bibitem{Gogoladze:2014cha}
  I.~Gogoladze, F.~Nasir, Q.~Shafi and C.~S.~Un,
  Phys.\ Rev.\ D {\bf 90} (2014) 3, 035008
  [arXiv:1403.2337 [hep-ph]].
\bibitem{Badziak:2014kea}
  M.~Badziak, Z.~Lalak, M.~Lewicki, M.~Olechowski and S.~Pokorski,
  JHEP {\bf 1503} (2015) 003
  [arXiv:1411.1450 [hep-ph]].
\bibitem{Kowalska:2015zja}
  K.~Kowalska, L.~Roszkowski, E.~M.~Sessolo and A.~J.~Williams,
  arXiv:1503.08219 [hep-ph].
\bibitem{Chakrabortty:2015ika}
  J.~Chakrabortty, A.~Choudhury and S.~Mondal,
  arXiv:1503.08703 [hep-ph].
\bibitem{Wang:2015rli}
  F.~Wang, W.~Wang and J.~M.~Yang,
  arXiv:1504.00505 [hep-ph].
\bibitem{Calibbi:2015kja}
  L.~Calibbi, I.~Galon, A.~Masiero, P.~Paradisi and Y.~Shadmi,
  arXiv:1502.07753 [hep-ph].
\bibitem{Davies:2011mp}
  R.~Davies, J.~March-Russell and M.~McCullough,
  JHEP {\bf 1104} (2011) 108
  [arXiv:1103.1647 [hep-ph]].
\bibitem{Abel:2009ve}
  S.~Abel, M.~J.~Dolan, J.~Jaeckel and V.~V.~Khoze,
  JHEP {\bf 0912} (2009) 001
  [arXiv:0910.2674 [hep-ph]].
\bibitem{moroi}
  T.~Moroi,
  Phys.\ Rev.\ D {\bf 53} (1996) 6565
  [Erratum-ibid.\ D {\bf 56} (1997) 4424]
  [hep-ph/9512396].
\bibitem{Graesser:2001ec}
  M.~Graesser and S.~D.~Thomas,
  Phys.\ Rev.\ D {\bf 65} (2002) 075012
  [hep-ph/0104254].
\bibitem{Chacko:2001xd}
  Z.~Chacko and G.~D.~Kribs,
  Phys.\ Rev.\ D {\bf 64} (2001) 075015
  [hep-ph/0104317].
\bibitem{Isidori:2007jw}
  G.~Isidori, F.~Mescia, P.~Paradisi and D.~Temes,
  Phys.\ Rev.\ D {\bf 75} (2007) 115019
  [hep-ph/0703035].
\bibitem{Kersten:2014xaa}
  J.~Kersten, J.-h.~Park, D.~St\"ockinger and L.~Velasco-Sevilla,
  JHEP {\bf 1408} (2014) 118
  [arXiv:1405.2972 [hep-ph]].
\bibitem{Chiu:2014oma}
  W.~C.~Chiu, C.~Q.~Geng and D.~Huang,
  Phys.\ Rev.\ D {\bf 91} (2015) 1,  013006
  [arXiv:1409.4198 [hep-ph]].
\bibitem{Adam:2013mnn}
  J.~Adam {\it et al.}  [MEG Collaboration],
  Phys.\ Rev.\ Lett.\  {\bf 110} (2013) 201801
  [arXiv:1303.0754 [hep-ex]].
\bibitem{HiggsMassLift}
  Y.~Okada, M.~Yamaguchi and T.~Yanagida,
  Prog.\ Theor.\ Phys.\  {\bf 85} (1991) 1;
  Phys.\ Lett.\ B {\bf 262} (1991) 54;
  J.~R.~Ellis, G.~Ridolfi and F.~Zwirner,
  Phys.\ Lett.\ B {\bf 257} (1991) 83;
  Phys.\ Lett.\ B {\bf 262} (1991) 477;
  H.~E.~Haber and R.~Hempfling,
  Phys.\ Rev.\ Lett.\  {\bf 66} (1991) 1815.
\bibitem{Brignole:2002bz}
  A.~Brignole, G.~Degrassi, P.~Slavich and F.~Zwirner,
  Nucl.\ Phys.\ B {\bf 643} (2002) 79
  [hep-ph/0206101].
\bibitem{Gunion:2002zf}
  J.~F.~Gunion and H.~E.~Haber,
  Phys.\ Rev.\ D {\bf 67} (2003) 075019
  [hep-ph/0207010].
\bibitem{Degrassi:2001yf}
  G.~Degrassi, P.~Slavich and F.~Zwirner,
  Nucl.\ Phys.\ B {\bf 611} (2001) 403
  [hep-ph/0105096].
\bibitem{Kitahara:2012pb}
  T.~Kitahara,
  JHEP {\bf 1211} (2012) 021
  [arXiv:1208.4792 [hep-ph]].
\bibitem{Carena:2012mw}
  M.~Carena, S.~Gori, I.~Low, N.~R.~Shah and C.~E.~M.~Wagner,
  JHEP {\bf 1302} (2013) 114
  [arXiv:1211.6136 [hep-ph]].
\bibitem{Kitahara:2013lfa}
  T.~Kitahara and T.~Yoshinaga,
  JHEP {\bf 1305} (2013) 035
  [arXiv:1303.0461 [hep-ph]].
\bibitem{Grothaus:2012js}
  P.~Grothaus, M.~Lindner and Y.~Takanishi,
  JHEP {\bf 1307} (2013) 094
  [arXiv:1207.4434 [hep-ph]].
\bibitem{Hisano:2010re}
  J.~Hisano and S.~Sugiyama,
  Phys.\ Lett.\ B {\bf 696} (2011) 92
   [Erratum-ibid.\ B {\bf 719} (2013) 472]
  [arXiv:1011.0260 [hep-ph]].
\bibitem{Martin:2002iu}
  S.~P.~Martin,
  Phys.\ Rev.\ D {\bf 66} (2002) 096001
  [hep-ph/0206136].
\bibitem{jaesmethod}
  J.-h.~Park,
  JCAP {\bf 1102} (2011) 023
  [arXiv:1011.4936 [hep-ph]];
  Phys.\ Rev.\ D {\bf 83} (2011) 055015
  [arXiv:1011.4939 [hep-ph]].
\end{thebibliography}
\end{document}